%% file: manuscript.tex
\documentclass[reqno]{amsart}

\usepackage[margin=3cm]{geometry}

\usepackage{arXiv_style}
\usepackage{fontawesome}

\usepackage[numbers,sort&compress]{natbib}
\usepackage{booktabs}
\usepackage[ruled,vlined]{algorithm2e}
\usepackage[most]{tcolorbox}
\usepackage{float}
\usepackage{url}
\usetikzlibrary{positioning,calc,arrows.meta}

\definecolor{greyYed}{RGB}{100,100,100}

\makeatletter
\def\blfootnote{\gdef\@thefnmark{}\@footnotetext}
\makeatother

\makeatletter
\renewcommand\paragraph{\@startsection{paragraph}{4}{\z@}{.5\linespacing\@plus.7\linespacing}{-.5em}{\normalfont\itshape}}
\renewcommand{\tocsection}[3]{\indentlabel{\@ifnotempty{#2}{\ignorespaces#2.\quad}}#3}
\makeatother

\newcommand{\doi}[1]{\href{https://doi.org/#1}{\textsc{doi}:\,\nolinkurl{#1}}}

\newcommand{\eraclitus}{\texttt{Eraclitus}\xspace}
\newcommand{\eraclitusfull}{\texttt{Eraclitus-4.7M}\xspace}
\newcommand{\rheon}{\textsc{Rheon}\xspace}

\newcommand{\MiniLMSix}{\texttt{all-MiniLM-L6-v2}}
\newcommand{\MiniLMTwelve}{\texttt{all-MiniLM-L12-v2}}
\newcommand{\MpnetBase}{\texttt{all-mpnet-base-v2}}
\newcommand{\MpnetPara}{\texttt{paraphrase-mpnet-base-v2}}

\tcbuselibrary{breakable}
\newcounter{questionCounter}
\newcommand{\Open}[1]{%
    \refstepcounter{questionCounter}%
    \begin{tcolorbox}[
        colframe=black!50,
        colback=gray!10,
        coltext=black,
        boxrule=0.8pt,
        before skip=4pt,
        after skip=4pt,
        top=4pt,
        bottom=4pt,
        left=6pt,
        breakable,
        right=6pt,
        boxsep=0.5pt
    ]
        \textbf{Open Question \thequestionCounter:}\xspace\emph{#1}
    \end{tcolorbox}
}

\newcounter{insightCounter}
\newcommand{\Insight}[1]{%
    \refstepcounter{insightCounter}%
    \begin{tcolorbox}[
        colframe=blue!50!black,
        colback=blue!5,
        coltext=black,
        boxrule=0.8pt,
        before skip=4pt,
        after skip=4pt,
        top=4pt,
        breakable,
        bottom=4pt,
        left=6pt,
        right=6pt,
        boxsep=0.5pt
    ]
        \textbf{Insight \theinsightCounter:}\xspace\emph{#1}
    \end{tcolorbox}
}

\tcbset{promptbox/.style={
    enhanced,
    coltitle=white, fonttitle=\bfseries\footnotesize,
    fontlower=\footnotesize\itshape,
    boxrule=0.7pt, arc=2pt, boxsep=1pt,
    left=5pt, right=5pt, top=3pt, bottom=3pt,
    before skip=5pt, after skip=5pt,
}}
\newtcolorbox{promptfalse}[1]{promptbox,
    colframe=redM, colback=redM!4, colbacklower=redM!10, title={#1}}
\newtcolorbox{prompttrue}[1]{promptbox,
    colframe=blueM, colback=blueM!4, title={#1}}

\title[Consensus and Factual Dynamics in LLM Populations]{Consensus and Factual Dynamics in Large Populations of Interacting Language Models}

\author[E.\ Ricco et al.]{}

\begin{document}

\blfootnote{$^{\star}$ Corresponding Author, (\href{mailto:emanuele.ricco@kaust.edu.sa}{\faEnvelopeO}) \texttt{emanuele.ricco@kaust.edu.sa}, (\href{https://emarich.github.io/}{\faGlobe}) \texttt{https://emarich.github.io/}}

\maketitle

\vspace{-1em}

\begin{center}
    \begin{minipage}{.89\linewidth}\centering
        \textsc{Emanuele Ricco}$^{\, \textsc{a},\star,}$,
        \textsc{Elia Onofri}$^{\, \textsc{a}, \orcidlink{0000-0001-8391-2563}}$,
        \textsc{Vincenzo Sammartino}$^{\, \textsc{a},\, \textsc{b}}$,
        \textsc{Roberto Di Pietro}$^{\, \textsc{a}}$.
        \\
        \bigskip
        \begin{minipage}{.9\linewidth}\centering
            \footnotesize
            $^\textsc{a}$Computer, Electrical and Mathematical Sciences and Engineering (CEMSE) Division,\\
            King Abdullah University of Science and Technology (KAUST)\\
            Thuwal 23955, Saudi Arabia
            \\[.5em]
            $^\textsc{b}$Dipartimento di Informatica, Universit\`a di Pisa\\
            Pisa, Italy
        \end{minipage}
    \end{minipage}
\end{center}

\medskip
\thispagestyle{empty}

\begin{abstract}

        Large Language Model (LLM) agents are increasingly deployed as populations of interacting entities, in which consensus --agreement on a shared answer-- emerges as a collective, unengineered behaviour.
        Prior work on LLM consensus shows that agents can cross-verify their answers and converge towards more factual responses, treating agreement as a proxy for correctness.
        However, these studies usually fix a single interaction structure, leaving open how consensus depends on how agents interact.
        We address this gap by introducing \rheon, a physics-inspired framework that recasts a population drawn from a single frozen model as an evolving $O(n)$ spin system on a ladder of interaction geometries of increasing effective dimension --from a 1D ring to a full-coupling mean-field graph-- with the sampling temperature $T$ as the tunable source of thermal disorder, evolved through a Glauber-like asynchronous dynamics.
        Sweeping \rheon across $432$ configurations of prompt, population size, communication topology, and sampling temperature yields \eraclitusfull, a tagged evolutionary corpus of $4.7$ million responses.
        We find that agents reach their strongest consensus gain within the first few update sweeps and that increasing the number of neighbours per agent accelerates convergence on average.
        We further show that whether a configuration settles on factually correct or hallucinated consensus is not predictable from its initial state alone, and that the hallucination-minimising temperature depends on how the agents are coupled, so the common near-greedy default is not automatically the safest.
        Finally, semantic agreement correlates positively with factual convergence, and interaction strengthens the association, yet never enough for unanimity to certify correctness.

    \medskip


\end{abstract}

\begin{multicols}{2}
    \tableofcontents
\end{multicols}

\newpage

\section{Introduction}\label{sec:introduction}

Multi-Agent Systems built from Large Language Models (LLMs) have moved quickly from research demonstrations to deployed infrastructure~\citep{wang2024survey,guo2024large}:
assistants that debate, negotiate, review one another's work, and vote on a shared answer are now common~\citep{wu2023autogen, li2023camel, Ma2025}.
Each constituent agent carries its own persona, role, or behavioural disposition, and it is the interaction among these agents (rather than any single one) that shapes how the population behaves as a whole.

This autonomy is precisely what makes the resulting systems hard to trust.
Once agents are left to interact without human oversight, there is no direct way to tell whether their collective output remains reliable:
a population can talk itself into a confident, unanimous, and wrong answer as readily as into a correct one~\citep{choi2025conformity, han2026conformity}.
Prior work on LLM consensus has largely treated agreement as a resource to be engineered --multi-agent debate drives agents towards a shared answer and, in doing so, tends to improve factuality~\citep{du2024improving,liang2024encouraging}-- reading convergence as a proxy for correctness~\citep{chuang2024simulating}.
Yet this leaves the governing variable unexamined:
how does the consensus a population reaches depend on the way its agents are wired together?

We argue that a population of interacting LLM agents is best viewed as a complex system, in which many distinct interaction geometries are possible and the pattern of interaction (rather than the individual agents) selects the collective outcome.
This reframes consensus as an emergent, macroscopic property of the population, and invites the tools statistical physics has long used to study ordering in systems of many coupled units:
order parameters, a control parameter playing the role of temperature, and a family of lattice geometries of increasing effective dimension.

Building on this view, we study the emergent behaviour of such populations along two coupled axes.
First, \emph{consensus}:
do agents that repeatedly rewrite their answers after reading their neighbours' converge on a single shared response, and how fast?
Second, \emph{factual dynamics}:
when they do converge, is that agreement predictably aligned with the truth, or can the same protocol manufacture a shared hallucination?
To answer these questions at scale we introduce \rheon --\emph{Response Homogenisation in Evolving $O(n)$ Networks}-- a framework that recasts each agent's response as two coupled spins: an $O(n)$ spin, the unit-normalised embedding of its text (capturing \emph{what} the agent says), and a binary factual spin recording \emph{whether} that response is correct or hallucinated.
The name follows Heraclitus' \emph{rh\'eon}, ``flowing'': a population of agent responses in perpetual flux, from which consensus nonetheless emerges.

Our key contributions are as follows.

\begin{itemize}

    \item We introduce \rheon, to the best of our knowledge the first statistical-mechanics-inspired framework for consensus analysis in evolving LLM agents.

    \item We instantiate this framework as a Glauber-inspired asynchronous dynamics and sweep it across a geometry ladder (1D ring, 2D torus, 3D torus, and mean-field), a range of population sizes, and sampling temperatures $T\in\{0.1,\dots,1.6\}$.

    \item Sweeping \rheon across all configurations, we build and release \eraclitusfull, a large-scale evolutionary dataset of LLM multi-agent interactions, entirely tagged for factuality under the LLM-as-a-judge paradigm.

    \item We analyse the evolutionary process under physics-inspired macroscopic order parameters to distil insights on how such populations behave and converge.
\end{itemize}

\section{Related Work}\label{sec:related_work}

Populations of LLM agents display collective behaviour that is not reducible to any single model, from believable social simulacra~\citep{park2023generative} to emergent interaction structure~\citep{demarzo2023emergence}.
This paper asks when such a population collectively orders, \ie agrees on one answer, or remains disordered, a question at the intersection of two literatures.

\paragraph{Statistical physics of LLMs.}
The vocabulary of statistical physics entered the LLM literature through the debate on \emph{emergent abilities} --capabilities reported to appear abruptly with scale~\citep{wei2022emergent}-- much of which Schaeffer et al.~\citep{schaeffer2023emergent} attribute to discontinuous metrics rather than to genuine discontinuities.
A more direct line applies thermodynamic diagnostics to generation itself:
Arnold et al.~\citep{arnold2024phase} locate transitions in the output distribution of LLMs as temperature and context vary, while Nakaishi et al.~\citep{nakaishi2024critical} and Toji et al.~\citep{toji2024bkt} report critical and Berezinskii--Kosterlitz--Thouless behaviour in the statistics of generated text.
These works probe a \emph{single} model's sampling distribution;
we instead borrow their descriptive apparatus --order parameters and a temperature-like control of disorder-- to characterise the collective dynamics of a \emph{population} of interacting agents.

\paragraph{Consensus.}
Consensus formation has long been studied in agent-based models of opinion dynamics: linear averaging~\citep{degroot1974reaching}, bounded-confidence dynamics~\citep{deffuant2000mixing,hegselmann2002opinion}, and the Ising-like and voter-model formulations reviewed by Castellano et al.~\citep{castellano2009statistical}, where lattice geometry and noise jointly determine whether global order is reachable;
a parallel control-theoretic tradition characterises consensus on networked multi-agent systems via graph connectivity~\citep{olfatisaber2007consensus}.
With LLMs, consensus has mainly been \emph{engineered}: multi-agent debate improves factuality and reasoning by driving agents towards agreement~\citep{du2024improving,liang2024encouraging}, and Chen et al.~\citep{chen2023multiagent} observe that negotiating LLM agents spontaneously adopt average-based strategies modulated by network topology.
A more recent line instead \emph{analyses} this behaviour, and finds it double-edged: Chuang et al.~\citep{chuang2024simulating} report a bias towards consensus on accurate information, whereas studies of \emph{conformity} show agents aligning with the numerical majority or the more confident peers even when they are wrong~\citep{choi2025conformity}; concurrently Han et al.~\citep{han2026conformity} find that denser topologies accelerate convergence at the cost of confident, incorrect cascades.
We build on this analytic line but differ on three fronts:
we recast the population as an evolving $O(n)$ spin system --our \rheon framework-- rather than an opinion-averaging model;
we sweep a \emph{geometry ladder} of increasing effective dimension (1D, 2D, 3D, and mean-field) jointly with population size and a genuine thermal control~$T$;
and we release the full evolutionary corpus it generates, \eraclitusfull, tagged for factuality.

\section{Methodology}\label{sec:methodology}
\subsection{The \rheon model}\label{sec:rheon}

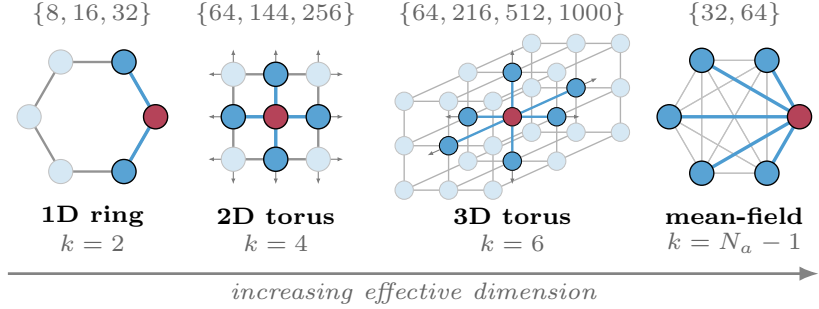
\begin{figure}[t]
    \centering
    \resizebox{0.7\columnwidth}{!}{\input{figs/geometry-ladder}}
    \caption{The geometry ladder.
    \rheon couples the $N_a$ agents through one of four graphs of increasing effective dimension:
    a 1D ring ($k=2$), a 2D torus ($k=4$), a 3D torus ($k=6$), and a mean-field all-to-all graph ($k=N_a-1$).
    In each panel one reference site (red) and its $k$ neighbours (blue) are highlighted, illustrating the coordination number;
    the population sizes $N_a$ used in the sweeps are reported above each panel.}
    \label{fig:geometry-ladder}
\end{figure}

Consider a population of $N_a$ agents, each an independent instance of the same frozen model (Qwen3-14B, 4-bit) asked to answer one fixed question~$p$.
The agents occupy the sites of a graph whose edges fix which of them may read one another, and we write $\mathcal N(i)$ for the neighbourhood of agent~$i$.
\rheon sweeps a \emph{geometry ladder} of four such graphs of increasing effective dimension (Figure~\ref{fig:geometry-ladder}):
a 1D ring, a 2D torus, and a 3D torus, each regular of coordination number $k=2\ell$ in lattice dimension $\ell$ under periodic boundary conditions, and a mean-field (MF) graph coupling every agent to all $N_a-1$ others.

The state of agent $i$ is its current textual response $r_i$, from which \rheon derives two coupled spins.
The \emph{semantic spin} $\mathbf s_i\in S^{d-1}$ is the whitened, unit-normalised sentence embedding of $r_i$ (Section~\ref{sec:order-parameters});
it records \emph{what} the agent says, and turns the population into a configuration of an $O(n)$ model whose dimension $n=d$ is fixed by the encoder.
The \emph{factual spin} $\sigma_i\in\{-1,+1\}$ is an Ising variable recording \emph{whether} the response is correct ($\sigma_i=-1$) or hallucinated ($\sigma_i=+1$), as adjudicated by the judge of Section~\ref{sec:eraclitus}.

Two competing forces act on this configuration:
peer interaction supplies an \emph{alignment coupling} that draws an agent's semantic spin towards its neighbours', while the generation \emph{temperature} $T$ plays the role of thermal disorder.
We disable nucleus and top-$k$ truncation ($\mathrm{top}_p=1$, $\mathrm{top}_k=0$) so that $T$ alone sets the entropy of the sampling distribution, and sweep $T\in\{0.1,0.4,0.7,1.0,1.3,1.6\}$ from the near-deterministic (cold) to the strongly disordered (hot) regime.

\paragraph{Dynamics.}
\rheon advances the configuration by a Glauber-like asynchronous dynamics (see Algorithm~\ref{algo:rheon} in Appendix~\ref{app:eraclitus}).
At initialisation ($t=0$) every agent answers the question independently, with no peer context, yielding a non-interacting reference population.
The system is then evolved for $\bar N=10$ sweeps.
A single sweep performs $N_a$ elementary updates;
each update draws an agent $i$ uniformly at random, shows it the \emph{current} responses of its neighbours $\mathcal N(i)$ together with its own previous answer, and overwrites $r_i$ with the model's reply.
Because neighbour context is read live, later updates within a sweep already see the effect of earlier ones --the sequential, asynchronous effect of Glauber dynamics-- rather than being applied in lockstep.

The dynamics therefore carries two distinct stochastic channels.
The sampling temperature randomises \emph{what} each agent writes, perturbing the token-level choices of every rewrite;
the random-sequential update order randomises \emph{when} each agent is refreshed, and hence which of its neighbours it has already seen move.

The random seed of each run is derived from its configuration, making the update sequence reproducible and each run resumable.
Every response is generated at a fixed length of exactly $100$ new tokens, which removes response length as a confounding variable.

\subsection{The \eraclitusfull dataset}\label{sec:eraclitus}

We obtain \eraclitusfull by sweeping \rheon across every configuration.
A \emph{configuration} is a tuple $(p, G, N_a, T)$ of prompt, geometry, size, and temperature, and one execution of the dynamics on a configuration is a \emph{run}.
We cross the six prompts described below with the four geometries at twelve sizes in total
--1D: $N_a\in\{8,16,32\}$;
2D: $N_a\in\{64,144,256\}$ ($L=8,12,16$);
3D: $N_a\in\{64,216,512,1000\}$ ($L=4,6,8,10$);
mean-field: $N_a\in\{32,64\}$--
and the six temperatures above, giving $6\times12\times6 = 432$ configurations.
Each configuration is run as $R=5$ independent trials, for $2{,}160$ runs in total;
throughout the paper, unless stated otherwise, every reported quantity is a mean $\pm$ standard deviation over these five trials.

One trial thus comprises $953{,}568$ LLM calls, dominated by the 3D lattice, and the full campaign $4{,}767{,}840$ calls --the $4.7$ million responses that name \eraclitusfull.

\paragraph{Prompts.}
The six prompts are multi-hop questions drawn from \textsc{FreshQA}~\citep{vu-etal-2024-freshllms} and listed in full in Appendix~\ref{app:eraclitus}.
Three carry a \emph{false premise} that the model must detect in order to answer correctly, and three are standard factual questions.
This split deliberately sets structural consensus --agents drawn towards their neighbours-- against semantic factuality --the correct answer-- so that ordering and correctness are free to diverge.
\begin{table}[t]
\centering
\scriptsize
\begin{tabular}{l r | r | r | r | r | r}
\hline
 & & \multicolumn{4}{|c|}{Cost per agent (m\$) through sweep $t$} & \\
\cline{3-6}
Geometry & $N_a$ & $t=0$ & $t=4$ & $t=7$ & $t=10$ & Total \\
\hline
1D         & 8    & 5.148 & 50.686  & 84.827  & 118.969  & \$0.95   \\
           & 16   & 5.148 & 50.689  & 84.821  & 118.935  & \$1.90   \\
           & 32   & 5.148 & 50.681  & 84.816  & 118.940  & \$3.81   \\
\hline
2D         & 64   & 5.148 & 66.308  & 112.153 & 157.992  & \$10.11  \\
           & 144  & 5.148 & 66.300  & 112.156 & 158.006  & \$22.75  \\
           & 256  & 5.148 & 66.305  & 112.156 & 157.999  & \$40.45  \\
\hline
3D         & 64   & 5.148 & 81.937  & 139.539 & 197.168  & \$12.62  \\
           & 216  & 5.148 & 81.957  & 139.595 & 197.271  & \$42.61  \\
           & 512  & 5.148 & 81.947  & 139.589 & 197.294  & \$101.01 \\
           & 1000 & 5.148 & 81.934  & 139.548 & 197.227  & \$197.23 \\
\hline
Mean-field & 32   & 5.148 & 279.404 & 485.222 & 691.095  & \$22.12  \\
           & 64   & 5.148 & 532.523 & 928.509 & 1324.774 & \$84.79  \\
\hline
\multicolumn{6}{l|}{\textbf{Total}} & \textbf{\$540.34} \\
\hline
\end{tabular}
\caption{Estimated cost of \eraclitus at Qwen3-14B per-token cloud pricing (\$0.10 / 1M input, \$0.24 / 1M output; \url{https://openrouter.ai/qwen/qwen3-14b}), broken down by configuration.}
\label{tab:cost-estimate}
\end{table}

Table~\ref{tab:cost-estimate} estimates the cost of generating \eraclitus at per-token cloud pricing;
the campaign itself --generation and tagging-- ran on two NVIDIA RTX A6000 GPUs (48\,GB, CUDA 12.2) over ${\sim}15$ days (see Appendix~\ref{app:eraclitus} for the call budget by geometry).
Note that across the three lattices the per-agent cost scales on average at $t=10$ from $119$\,m\$ (1D) to $158$\,m\$ (2D) to $197$\,m\$ (3D) as each agent reads $2$, $4$ and $6$ peers.
However, in mean-field a single update feeds an agent all $N_a-1$ responses, and truncating at the $t=4$ plateau it costs $532$\,m\$ per agent at $N_a=64$, for a total of \$$34.08$ --below the full-campaign cost of 2D at $N_a=256$ (\$$40.45$) and of 3D at every size above $N_a=64$.

\paragraph{Factual tagging.}
To relate semantic ordering to factual correctness, every response of \eraclitus is tagged by an LLM-as-a-judge (Qwen2.5-72B-Instruct) into one of four categories:
\emph{correct}, \emph{hallucinated}, \emph{not-known} (the agent explicitly declines to answer), and \emph{judge-failure} (the judge returns no parseable verdict).
The factual analysis retains the correct and hallucinated responses, mapping them to the factual spin $\sigma_i$ of Section~\ref{sec:rheon}, and discards the two residual categories.
A stratified sample of the judge's verdicts is validated against human annotators;
the tagging prompt, the human-validation protocol, and a full description of \eraclitus are given in Appendices~\ref{app:eraclitus} and~\ref{app:human-validation}.
The tags are released together with the response corpus as part of \eraclitus.

\subsection{Order parameters}\label{sec:order-parameters}

We analyse each run along two coupled behavioural axes, consensus and factuality, defining the macroscopic observables below on the whitened spins.

\paragraph{Embedding and whitening.}
Each response is embedded with \MiniLMSix{} into a raw vector $\mathbf e\in\mathbb R^{d}$ ($d=384$);
three further sentence encoders are used only for a robustness check reported in Appendix~\ref{app:embedding-robustness}.

Raw sentence-transformer embeddings are strongly anisotropic, occupying a narrow cone in which even unrelated sentences share a large positive cosine similarity, so that a naive overlap $\mathbf e_i\cdot\mathbf e_j$ is dominated by this common component rather than by genuine agreement~\citep{whitening2021}.
We remove this bias by whitening, once per prompt.
From that prompt's independent $t=0$ responses --pooled across all temperatures, sizes, and geometries-- we estimate a mean $\boldsymbol\mu$ and a per-dimension standard deviation $\boldsymbol\varsigma$, and map each raw embedding to $\tilde{\mathbf e} = (\mathbf e - \boldsymbol\mu)\oslash\boldsymbol\varsigma$ with $\mathbf s = \tilde{\mathbf e}\,/\,\lVert\tilde{\mathbf e}\rVert$,
where $\oslash$ denotes the element-wise (Hadamard) quotient.
The whitened spins $\mathbf s\in S^{d-1}$ populate an approximately isotropic space in which the dot product $\mathbf s_i\cdot\mathbf s_j$ measures genuine semantic alignment and the order parameters below are meaningful.

\paragraph{Semantic consensus.}
The primary measure of semantic order is the mean resultant vector of the spins, the \emph{semantic magnetisation}, and its magnitude:
\begin{equation}
\mathbf{m}(t) = \frac1{N_a}\sum_{i=1}^{N_a}\mathbf{s}_i(t),
\qquad
m(t) = \lVert \mathbf{m}(t)\rVert \in [0,1]\ .
\end{equation}
Here $\mathbf m(t)$ is the average of the unit spins --not a centroid in the raw embedding space-- and its magnitude $m(t)$ is the standard $O(n)$ order parameter; Figure~\ref{fig:order-schematic} in Appendix~\ref{app:order-parameter} illustrates it schematically.
When every agent points in the same direction the spins add coherently and $m=1$, a monolithic semantic consensus;
as they spread over the sphere their contributions cancel and $m$ falls.
For $N_a$ independent, isotropically oriented spins $m$ does not vanish but fluctuates around a finite-size floor of order $1/\sqrt{N_a}$, the disordered baseline;
intermediate values describe a partially ordered population, a single loose cluster or a majority direction coexisting with a dissenting minority.

Because $m$ carries this $1/\sqrt{N_a}$ floor, we also track its intensive, floor-free counterpart, the \emph{semantic cohesion}
\begin{equation}
\bar{c}(t) = \frac{2}{N_a(N_a-1)}\sum_{i<j}\mathbf{s}_i(t)\cdot\mathbf{s}_j(t)\ .
\end{equation}
This mean pairwise overlap carries the same information as $m$ at fixed size --$\bar c = (N_a m^2-1)/(N_a-1)$-- but subtracts the finite-size floor, vanishing for a disordered population at every $N_a$ and hence comparable across sizes.

\paragraph{Factual dynamics.}
The consensus observables track whether agents agree, not whether they are right.
Using the factual spin $\sigma_i(t)$, we summarise the population's factual state by the \emph{hallucination density}
\begin{equation}
\rho(t) = \frac1{N_a}\sum_{i=1}^{N_a}\frac{1+\sigma_i(t)}{2}\in[0,1]\ ,
\end{equation}
\ie the fraction of the population holding a wrong answer at time $t$.
Read alongside $m$ and $\bar c$, it separates the two ways a population can reach consensus, as Figure~\ref{fig:factual-outcomes} in Appendix~\ref{app:order-parameter} illustrates schematically:
$\rho\to0$ at high cohesion is collective error correction, whereas $\rho\to1$ at high cohesion is a shared hallucination.
The effect of interaction alone is isolated by the signed difference $\Delta\rho=\rho(t{=}10)-\rho(t{=}0)\in[-1,1]$, negative when interaction corrects errors, positive when it manufactures them, and zero under a symmetric voter model.

\section{Results}\label{sec:results}

We now report how the three order parameters evolve as the population runs across the geometry ladder and the temperature sweep.
Analyses that follow a single prompt in detail use Prompt~3 for the trajectory panels --its dynamics are the clearest to read-- and the cleanly-tagged Prompt~6 for the temperature analysis, whose factual optimum needs reliable tags.
Analyses that pool across prompts use prompts~2--6 and exclude Prompt~1, whose false premise misleads the judge by tagging the correct refutations as hallucinated and the confident wrong answers as correct, agreeing with our reference annotator on only $34\%$ of the audited responses ($\kappa=0.025$) against a $98.58\%$ average across both annotators on prompts~2--6;
validation protocol, per-prompt agreement table, and robustness checks under alternative embedding models are provided in Appendices~\ref{app:human-validation} and~\ref{app:embedding-robustness}.
The section is organised around four questions, answered directly in the upcoming Discussion.

\Open{How does the interaction geometry, \ie the ladder from a 1D ring to full mean-field coupling, shape whether, how fast, and how completely a population reaches consensus?}

\Open{Is the factual fate of a population, \ie whether it repairs error or settles on a shared hallucination, fixed by its initial accuracy, or can a correct minority overturn a hallucinated majority?}

\Open{Does the sampling temperature act as a genuine control on the factual outcome, and if so, is there a single temperature that minimises hallucination --the common near-greedy default, perhaps-- or does the safest setting depend on how the agents are coupled?}

\Open{Is the population's consensus reached slowly or is it front-loaded; and if the outcome cannot be predicted in advance, can it at least be read off early?}

\begin{figure*}[t]
    \centering
    \includegraphics[width=.9\linewidth]{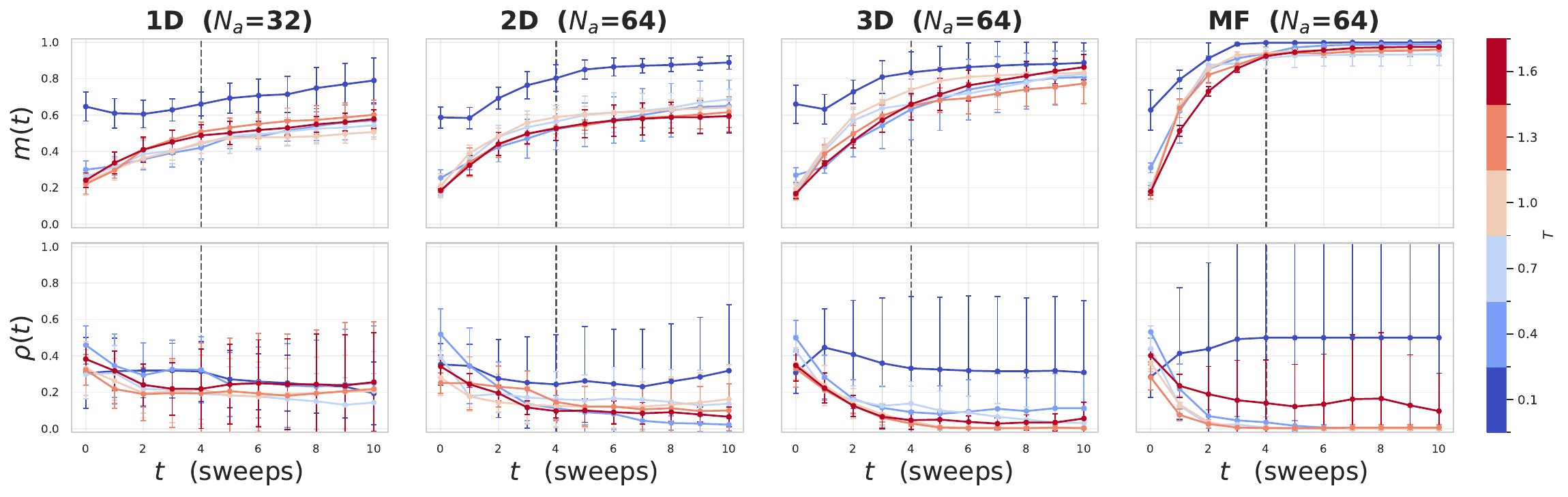}
    \caption{Temporal evolution of the semantic magnetisation $m(t)$ (top) and of the hallucination density $\rho(t)$ (bottom) across network geometries for Prompt 3, shown for the 1D ring at $N_a=32$ and for 2D, 3D, and MF at $N_a=64$, over ten sweeps.
    Trajectories are colour-coded by sampling temperature $T$.}
    \label{fig:evolution}
\end{figure*}

\subsection{Single-prompt dynamics}\label{sec:results-single}

\begin{figure}[t]
    \centering
    \captionsetup{margin=0pt}
    \begin{minipage}[t]{.49\linewidth}\centering
        \includegraphics[width=\linewidth]{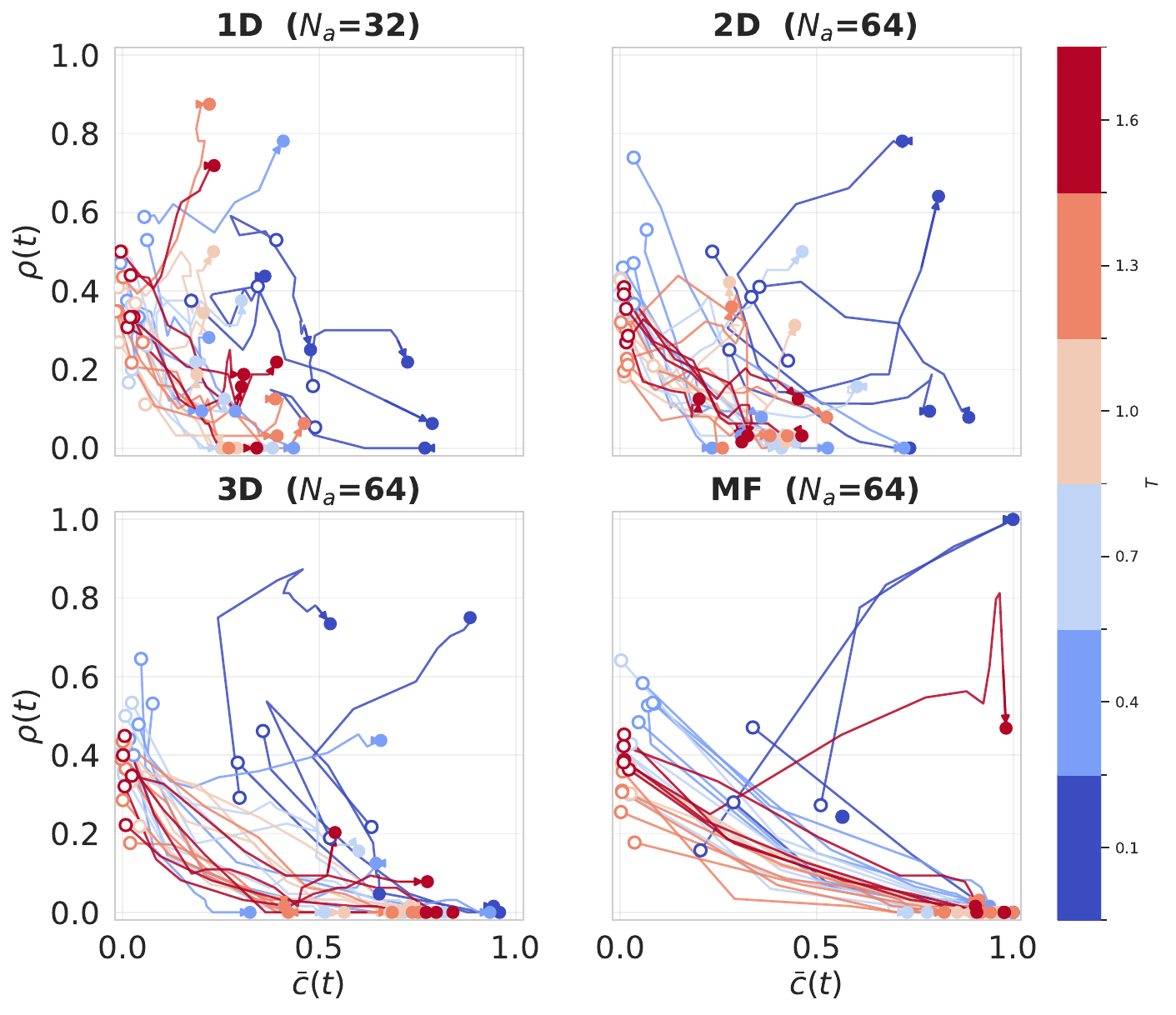}
        \caption{Joint evolution of hallucination density $\rho(t)$ against linguistic consensus $\bar{c}(t)$ for Prompt 3.}
        \label{fig:phase_evolution}
    \end{minipage}\hfill
    \begin{minipage}[t]{.47\linewidth}\centering
        \includegraphics[width=\linewidth]{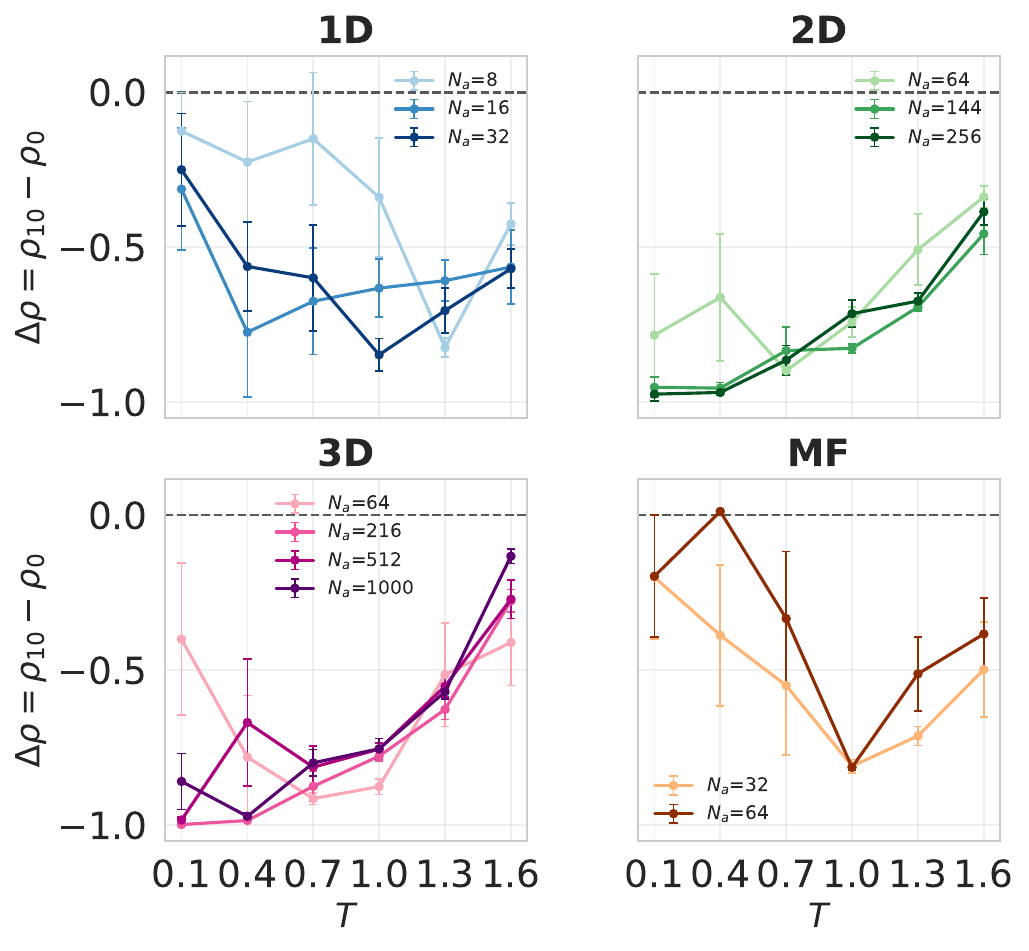}
        \caption{Interaction-induced change in hallucination, $\Delta\rho = \rho(t{=}10)-\rho(t{=}0)$, against sampling temperature $T$ for Prompt~6, resolved by geometry and system size.}
        \label{fig:temperature_by_geometry}
    \end{minipage}
\end{figure}

Figures~\ref{fig:evolution} and~\ref{fig:phase_evolution} follow Prompt~3 up the geometry ladder, showing the tension between consensus and factuality.
Panels are matched at $N_a=64$ (except the 1D ring at $N_a=32$) so that differences reflect geometry rather than system size.

Figure~\ref{fig:evolution} (top) shows that every geometry orders.
The magnetisation rises sharply over the first sweeps and has flattened by $t\approx4$, as has the factual curve beneath it; the level at which it settles, however, is set by the geometry:
the ring and the 2D torus saturate well short of full alignment, the 3D torus climbs higher, and the mean-field graph reaches near-perfect alignment ($m\approx1$) at every temperature.
At $T=0.1$ the agents already start half-aligned ($m(0)\approx0.63$, cf.\ $\approx0.2$ at $T=1.6$), near-greedy sampling having made independent responses nearly identical before any interaction took place, making colder curves sit higher throughout:
this is why we pose the temperature question on the factual axis, where the initial level can be differenced out.

Ordering, however, carries no guarantee of correctness (Figure~\ref{fig:evolution}, bottom): a population can settle just as firmly on a wrong answer as on the right one.
Denser connectivity generally aids error correction, but neither uniformly across temperatures nor as a guarantee of recovery.
In the MF limit, the dense web of inter-agent feedback drives $\rho(t)\to0$ within the first $t=4$ sweeps at all but the coldest temperature.
At $T=0.1$, the same connectivity instead freezes the population into a single absorbing consensus, and which consensus it reaches varies across trials: from full recovery ($\rho\to0$) to a shared hallucination ($\rho\to1$), so the pooled hallucination density \emph{rises} from ${\approx}0.29$ to ${\approx}0.50$.

What the coldest setting does on every geometry denser than the ring is therefore not to raise the hallucination density smoothly but to make the outcome \emph{bimodal}, and the trial-to-trial spread this produces is as large as the effect itself ($\rho(10)=0.32\pm0.36$ in 2D, $0.31\pm0.40$ in 3D, $0.50\pm0.58$ in MF over $R=5$).
Two features survive that dispersion.
The ring never locks in at all --its coldest trials end between $\rho=0$ and $0.44$, its pooled endpoint $\rho\approx0.19$ among the lowest curves of its own panel-- whereas each denser geometry locks in on two of its five trials;
and the \emph{depth} of a lock-in does track the coordination number, the 2D and 3D ones settling around $\rho\approx0.6$--$0.8$ while both mean-field lock-ins are total ($\rho=1$, every agent on the same wrong answer).
Sparser coupling thus protects the agents from freezing outright, at the cost of a higher residual hallucination floor retained at every other temperature ($\rho\approx0.14$--$0.26$ on the ring at $N_a=32$).

Read against each other (see Figure~\ref{fig:phase_evolution}), the two axes explain why cohesion cannot stand in for correctness.
The trajectories sweep from the top-left --disordered and factually mixed-- toward the bottom-right, so agreement and repair typically arrive together, consistent with the weakly positive cohesion--accuracy correlations reported below.
The cohesion a population can reach climbs the ladder much as the magnetisation ceiling does, approaching $\bar c\approx1$ only in MF.
The informative feature, though, is the discordant branch.
On every geometry denser than the ring, the coldest trajectories peel off to the top-right instead:
in MF at $T=0.1$ the population arrives at $\bar c\approx1$ with $\rho\approx1$, a state where every agent gives the same confident wrong answer.
A population can hence be at its most cohesive exactly where it is least reliable: dense coupling is what makes the wrong corner reachable, and cold sampling is what steers into it.
Whether temperature is a systematic control on that choice is what we test next.

\smallskip

Figure~\ref{fig:temperature_by_geometry} asks whether sampling temperature is a genuine control on factual outcome, using the cleanly-tagged Prompt~6, whose agents begin in a shared factual error (misattributing the Perceptron) that interaction may or may not repair.
At its best $T$, interaction removes most of that error in every geometry:
the deepest change reaches $\Delta\rho=-0.78$ to $-1.00$ depending on geometry and size, against the $\Delta\rho=0$ expected of a symmetric voter model.

Which temperature repairs best, however, shifts with the coupling.
The larger 2D and 3D lattices repair most deeply at (or near) the coldest setting ($\Delta\rho=-0.98$ at $T=0.1$ for 2D at $N_a=256$ and $-1.00$ for 3D at $N_a=216$), the correct minority propagating through many neighbours without needing noise;
MF instead repairs best at an intermediate $T^\star\approx1.0$ at both of its sizes, its cold settings barely repairing at all ($-0.20$ at $T=0.1$ and $+0.01$ at $T=0.4$ for $N_a=64$, against $-0.81$ at $T^\star$).
The smallest lattices side with mean-field rather than with their own geometry ($T^\star=0.7$ at $N_a=64$ in both), so the cold optimum needs a large local neighbourhood \emph{and} a population big enough to sustain a correct cluster within it.
The near-greedy default is therefore not uniformly safe:
where the coupling is global rather than local it freezes the population in its initial (potentially wrong) consensus, and some sampling noise is what lets the correct minority escape it.
Note that this ranking is a property of the coupling and not of the headroom each temperature starts from:
at $t=0$ the agents have not yet read one another, so $\rho(t{=}0)$ is fixed by the prompt and $T$ alone, identical across geometries, and the fixed-prompt comparison across geometries is clean by construction.
Read against the Prompt~3 panels above, this sharpens rather than contradicts them.
The two prompts begin at opposite ends of the factual axis, and $T$ moves that starting point --raising the initial error where a prompt sits near the accuracy ceiling, lowering it where the population starts uniformly wrong-- so their raw $\Delta\rho$ curves record different headroom and are not directly comparable.
Measured instead as the share of the initial error that interaction removes, the two agree: repair is weakest at both ends of the sweep and strongest between them.
What temperature buys is therefore not prompt-specific:
the hottest setting is never the best one, and how far towards the cold end the optimum moves is governed by the coupling, as later discussed in Section~\ref{sec:discussion}.

\subsection{Aggregate behaviour across prompts}\label{sec:results-pooled}

\begin{figure}[t]
    \centering
    \includegraphics[width=0.8\linewidth]{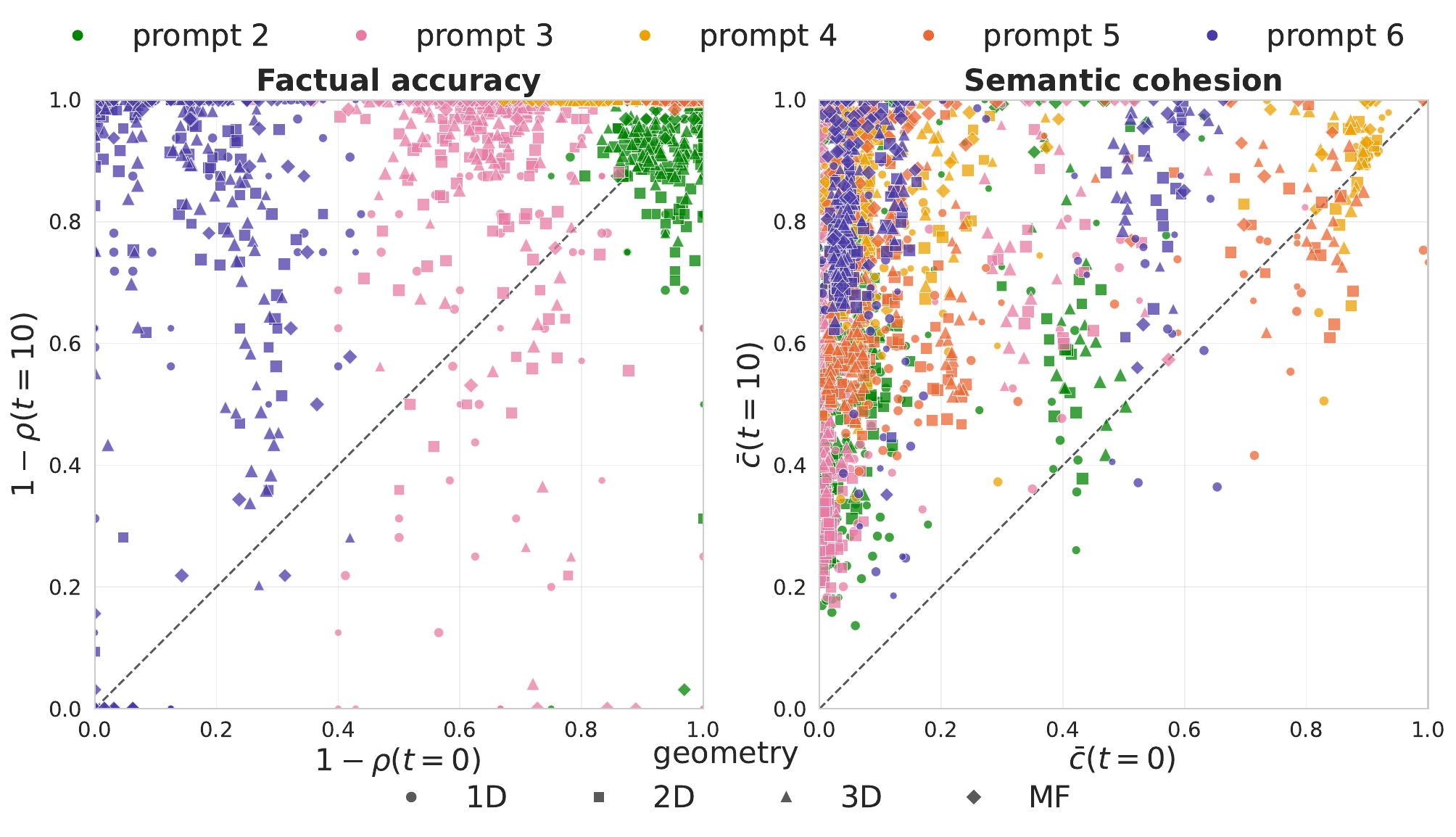}
    \caption{Global transformation from $t=0$ ($x$-axis) to $t=10$ ($y$-axis) of factual accuracy (left, $1-\rho$) and semantic cohesion (right, $\bar c$) divided by prompt (2--6, colour) and network geometry (shape).
    The invariant threshold ($y=x$ line) is shown as a dashed line, with the upper and lower triangles indicating improved and worsened performance over time, respectively.}
    \label{fig:accuracy_cohesion}
\end{figure}

The trajectories above follow one prompt at a time, so the factual axis they trace inherits whatever that prompt happens to ask.
We therefore turn to the pooled outcomes, treating each run as a single point --where it starts and where it ends on each axis-- rather than as a trajectory (Figure~\ref{fig:accuracy_cohesion}).

With the unreliable Prompt~1 excluded, the accuracy panel is repair-dominant.
The two prompts that begin with substantial error --Prompts~3 and~6-- move decisively above the $y=x$ line toward higher accuracy, the near-saturated Prompts~4 and~5 stay high on it, and Prompt~2 scatters about it with a negligible mean displacement.
No prompt's cluster sits below the diagonal on average, but amplification does survive at the level of individual runs:
about $5\%$ of those pooled here ($85$ of $1{,}800$) have $\Delta\rho>0.1$, and half of these sit on the 1D ring --the sparsest coupling in the ladder-- against only four in mean-field, consistent with the residual hallucination floor that sparse coupling retains.
That a corrupted initial majority is nonetheless routinely pulled back toward the correct answer indicates that the dynamics is not a symmetric voter model~\citep{clifford1973model,holley1975ergodic}, in which the initial majority fraction would instead be preserved in expectation.

The cohesion panel (Figure~\ref{fig:accuracy_cohesion}, right) shows a markedly different picture:
essentially every point, across prompts~2--6 and all four geometries, lies above the identity line.
Regardless of the interaction structure, the population moves from a low- or moderate-cohesion state at $t=0$ toward substantially higher cohesion by $t=10$;
geometry and prompt here shape \emph{how far} and \emph{how quickly} cohesion rises, but essentially never whether it rises at all.
Consensus formation is therefore a near-universal outcome of the interaction protocol, while the factual correctness of that consensus is not.

Table~\ref{tab:cohesion-accuracy-corr-size} quantifies this relationship directly, reporting the Pearson correlation between cohesion $\bar c$ and accuracy $1-\rho$ resolved by geometry and system size, pooled over prompts~2--6, temperatures, and trials, at $t=0$ and at $t=10$.

\begin{table}[t]
\centering
\small
\begin{tabular}{ccll}
\toprule
    & $N_a$ & $t=0$ & $t=10$ \\
\midrule
\multirow{3}{*}{1D}\hspace{-.35em}
    & 8 & +.127 & +.183$^\star$ \\
    & 16 & +.050 & +.024 \\
    & 32 & +.068 & +.237$^\star$ \\
\hline
\multirow{3}{*}{2D}\hspace{-.35em}
    & 64 & +.066 & +.162$^\star$ \\
    & 144 & +.095 & +.327$^\star$ \\
    & 256 & +.097 & +.361$^\star$ \\
\bottomrule
\end{tabular}
\begin{tabular}{ccll}
\toprule
    & $N_a$ & $t=0$ & $t=10$ \\
\midrule
\multirow{4}{*}{3D}\hspace{-.35em}
    & 64 & +.084 & +.128 \\
    & 216 & +.094 & +.144 \\
    & 512 & +.111 & -.029 \\
    & 1000 & +.118 & +.060 \\
\hline
\multirow{2}{*}{MF}\hspace{-.35em}
    & 32 & +.084 & +.230$^\star$ \\
    & 64 & +.091 & +.030 \\
\bottomrule
\end{tabular}
\caption{Pearson correlation over 150 runs between semantic cohesion $\bar c$ and factual accuracy ($1-\rho$) evaluated before ($t=0$) and after ($t=10$) the agents' evolution. $^\star = p<0.05$.}
\label{tab:cohesion-accuracy-corr-size}
\end{table}

\section{Discussion}\label{sec:discussion}

We now return to the four questions opened in Section~\ref{sec:results}, stating each answer as an insight and discussing what it does --and does not-- license.

\Insight{Consensus is a near-universal outcome, ordering on every geometry, and this happens faster and more completely the higher the connectivity.}

Consensus is not surprising because an agent that rewrites its answer is influenced towards its neighbours with nothing that pushes back.
The interesting behaviour is in the ladder: sparse coupling lets locally ordered domains survive side by side, whereas global coupling typically admits only one, and the magnetisation and cohesion ceilings climb accordingly.
Mean field reaches near-perfect alignment fastest and at every $T$, while also being the most expensive per agent (Table~\ref{tab:cost-estimate}).

\Insight{A hallucinated majority is not a stable verdict:
a correct minority can pull the population back to the right answer, so a configuration's factual outcome is not fixed by its initial vote, \ie the dynamics is not a symmetric voter model.}

A population does not merely ratify its initial majority:
what we observe is a systematic, often near-complete drift towards the genuine answer, which is not interchangeable with a hallucinated one.
Repair is not a guarantee --neighbours already stating the same wrong answer can amplify it-- so we read repair as the general behaviour and the shared hallucination as its narrow, identifiable failure mode rather than a symmetric alternative.

\Insight{Sampling temperature is a genuine control on the factual outcome and not a monotone one: where the optimum falls is set by how the agents are coupled.}

One caveat governs how the sweep must be read: $T$ acts \emph{before} any interaction takes place, since near-greedy sampling already makes independent responses nearly identical, so a comparison across temperatures is a comparison of the \emph{change} each induces and never of the level it starts from.
Differenced that way, the hottest sampling is never the safest, and the coldest is safest only where the coupling is local and the population large enough to sustain a correct cluster within a neighbourhood; under global coupling the optimum lies inside the sweep, at $T^\star\approx1.0$.
A single mechanism covers both ends.
Low temperature commits each agent to the most probable continuation of its context, and therefore amplifies whatever that context already carries:
under local coupling a correct cluster need outweigh only a few neighbours, so cold sampling drives the repair home;
under global coupling every agent reads the whole population, so the same commitment locks in the majority it began with, and it is then sampling noise that gives a correct minority any room to overturn it, much as a stochastic optimiser relies on noise to leave local minima.
As such, $T$ must be matched to the interaction.

\Insight{Both consensus and factuality are front-loaded, settling within the first few sweeps, with the first sweeps already making a population's outcome legible.}

Front-loading is what licenses truncating a deployed run at $t=4$, where both curves have flattened (Figure~\ref{fig:evolution}):
it removes $6$ of the $11$ calls each agent would otherwise make ($54.5\%$) --at our campaign's scale, the difference between \$$540$ and \$$223$ (Table~\ref{tab:cost-estimate})-- at no measurable cost in either observable.
The same trace supports a claim about monitoring rather than about prediction:
the previous insight denies any forecast of the factual outcome from the initial responses, yet that outcome is legible from the first few sweeps of the trajectory itself --the sweeps in which the compute is still cheap.
A deployed population can therefore be watched rather than trusted, yet it must not be monitored with agreement alone:
cohesion and accuracy do correlate positively in every geometry, and interaction strengthens the association (Table~\ref{tab:cohesion-accuracy-corr-size}, mean $r$ rising from $+0.09$ to $+0.15$), but $r$ peaks at $+0.36$, half the configurations do not reach significance, and among the most cohesive states we observe are the confidently wrong ones.

\paragraph{Scope and limitations.}
The population is homogeneous by construction --one frozen backbone, one prompt per run, no personas or adversaries-- on regular geometries, so we characterise a single model rather than a heterogeneous deployment.
The factual axis rests on automatic tags that are noisiest where a question carries a false premise, so $\rho$ is a coarse signal and the judge-independent $m$ and $\bar c$ carry the consensus claims.
Every statement about front-loading, and about apparently absorbing states, holds within a horizon of ten sweeps.

\section{Conclusion}\label{sec:conclusion}

We have introduced \rheon, a framework that recasts a population of interacting LLM agents as an evolving $O(n)$ spin system, and swept it across $432$ configurations of prompt, geometry, population size, and sampling temperature to produce \eraclitusfull, a corpus of $4.7$ million factually tagged responses.
Across that sweep consensus is the generic outcome of the protocol, settled within the first few sweeps and bounded in completeness by the interaction geometry, whereas the factual state it settles on is not fixed by the initial majority and repair remains available even when hallucinations dominate initially;
sampling temperature controls that outcome only in a way set by the coupling, so the near-greedy default common in deployment is not automatically the prudent one.
Two directions follow: annealing $T$ over the sweep, to buy escape from a wrong consensus early and stability late; and relaxing the homogeneity noted above, to test whether the near-universal ordering survives once the agents are no longer interchangeable.

All code needed to reproduce the generation, the observables, and every figure will be made publicly available, together with \eraclitusfull and its judge tags, under a permissive research licence.

\bibliographystyle{elsarticle-num}
\bibliography{arXiv_biblio}

\newpage

\appendix

\noindent The appendices contain the following information not reported in the main text:
\begin{itemize}
    \item The full description of the \eraclitus dataset: the six benchmark prompts; the agent prompt template; the generation parameters; the LLM-as-a-judge tagging protocol; the pseudocode for the \rheon dynamics; the file layout and row schema of the released corpus; and the LLM call budget.
    \item A breakdown of the judge's tagging results by geometry, temperature, and prompt.
    \item The human-validation protocol and its results.
    \item A geometric reading of the order parameters, and the full plots of every observable shown in the main text, extended to the five analysed prompts (Prompts~2--6).
    \item The embedding-robustness results.
\end{itemize}

\section{The \eraclitusfull dataset}\label{app:eraclitus}
In this section we describe in detail the \eraclitusfull dataset, which is released alongside this paper.

\eraclitusfull is built from six fixed, never-changing questions drawn from \textsc{FreshQA}~\citep{vu-etal-2024-freshllms}.
The first three carry a \emph{false premise} that the model must detect in order to answer correctly;
the second three are standard multi-hop factual questions with no false premise.
This split sets structural consensus (agents drawn toward their neighbours) against semantic factuality (the correct answer), so that ordering and correctness can diverge.

\subsection{The prompt list}

\paragraph{False premise (the system must detect the fallacy).}
The lower half of each box states the premise the agent has to reject in order to answer correctly.

\begin{promptfalse}{Prompt 1}
``What is the smallest cube number which can be expressed as the sum of two different positive cube numbers in two different ways?''
\tcblower
By Fermat's Last Theorem, no cube is the sum of two positive cubes.
\end{promptfalse}

\begin{promptfalse}{Prompt 2}
``By how many basis points did the Federal Reserve cut interest rates from August to December 2022?''
\tcblower
The Fed raised rates during this period; it did not cut them.
\end{promptfalse}

\begin{promptfalse}{Prompt 3}
``How much longer will Brittney Griner spend in the Russian prison?''
\tcblower
Brittney Griner was released in December 2022.
\end{promptfalse}

\paragraph{No false premise (standard multi-hop factual).}
These three carry no false premise: the question is answerable as asked.

\begin{prompttrue}{Prompt 4}
``How old was Queen Elizabeth II when she died?''
\end{prompttrue}

\begin{prompttrue}{Prompt 5}
``After which country did Marie Curie name the first element that she discovered?''
\end{prompttrue}

\begin{prompttrue}{Prompt 6}
``What killed the student inventor of the Perceptron?''
\end{prompttrue}

\subsection{The agent prompt template}
Every LLM call is built from one chat template, tuned along three axes:
\textbf{time} $t$ (context present or not), \textbf{temperature} $T$ (sampling parameter), and \textbf{neighbour count} $k$ (set by geometry degree).

\emph{$t = 0$ --- independent generation (no context).}
\begin{quote}
\ttfamily
{[}\\
\hspace*{1em}\{"role": "system",\\
\hspace*{2em}"content": "Answer the BASE\_PROMPT."\},\\
\hspace*{1em}\{"role": "user",\\
\hspace*{2em}"content": \{"BASE\_PROMPT": "<FreshQA prompt $P_p$>"\}\}\\
{]}
\end{quote}

\emph{$t \geq 1$ --- peer-conditioned update ($k$ neighbours).}
\begin{quote}
\ttfamily
{[}\\
\hspace*{1em}\{"role": "system",\\
\hspace*{2em}"content": "Answer the BASE\_PROMPT considering your peers'\\
\hspace*{2em}NEIGHBOR\_RESPONSES and YOUR\_PREVIOUS\_RESPONSE."\},\\
\hspace*{1em}\{"role": "user",\\
\hspace*{2em}"content": \{\\
\hspace*{3em}"BASE\_PROMPT": "<FreshQA prompt $P_p$>",\\
\hspace*{3em}"YOUR\_PREVIOUS\_RESPONSE": "<$r_i$ at $t-1$>",\\
\hspace*{3em}"NEIGHBOR\_RESPONSES": \{"Neighbor\_1": "...",\\
\hspace*{6em}"...": "...",\\
\hspace*{6em}"Neighbor\_k": "..."\}\}\}\\
{]}
\end{quote}

\subsection{The generation parameters}
All agents were instantiated under Qwen3-14B, 4-bit quantisation, thinking mode disabled.

To standardise the output, we adopted a fixed response length, setting \texttt{max\_new\_tokens} $=$ \texttt{min\_new\_tokens} $= 100$:
this way, the end-of-sequence token is suppressed until the $100$-token floor is reached, so every response is exactly $100$ tokens long.

Finally, we set $\mathrm{top}_p = 1$, $\mathrm{top}_k = 0$ making temperature the only relevant sampling parameter.

\subsection{The factuality tagging protocol}
Every stored response is classified by an independent LLM-as-a-judge, \texttt{Qwen2.5-72B-Instruct}, queried once per response with the question text and the response text substituted into the template below (\texttt{<prompt>} and \texttt{<response>} are the substitution slots):

\begin{quote}
\ttfamily\raggedright\small
Analyze this response and classify it into
one of three categories.\\[4pt]
QUESTION: <prompt>\\[4pt]
RESPONSE: <response>\\[4pt]
Classify the response as:\\
- 0: CORRECT (factually accurate)\\
- 1: HALLUCINATED (contains false\\
\hspace*{2em}claims, contradictions, fabricated\\
\hspace*{2em}information or does not answer the
\hspace*{2em}question)\\
- 2: NOT\_KNOWN (model admits it doesn't\\
\hspace*{2em}know)\\[4pt]
Provide your verdict:\\
CATEGORY: [0, 1, or 2]
\end{quote}

\noindent The integer verdict is parsed from the judge's output into the three categories \emph{correct} ($0$), \emph{hallucinated} ($1$), and \emph{not-known} ($2$);
responses for which no integer can be parsed are assigned a fourth category, \emph{judge-failure} ($-1$).
The factual analyses in the main text retain the \emph{correct} and \emph{hallucinated} rows and discard \emph{not-known} and \emph{judge-failure}. Furthermore, a more detailed description of the human validation of the judge is given in Section~\ref{app:human-validation} while statistics for the tagging are provided in Section~\ref{sec:llm-tagging}.

\paragraph{A failure mode of the judge: Prompt 1.}\label{par:prompt1-caveat}
The judge inherits the false premise of Prompt~1.
By Fermat's Last Theorem no cube is the sum of two positive cubes, so the correct response is a refutation of the question;
the judge instead treats the taxicab number $1729$ --which is a sum of two positive cubes in two ways, but is not itself a cube-- as the expected answer.
The two adjudicated labels are therefore not merely noisy on this prompt but systematically \emph{inverted}:
the judge tags the confident $1729$ answers \emph{correct} and the correct refutations \emph{hallucinated}, as quantified in Section~\ref{app:human-validation}.
Prompt~1 is consequently excluded from every pooled analysis in the main text.
Its tagged counts are still reported in Section~\ref{sec:llm-tagging}, since the responses themselves ship with the dataset, but every Prompt-1 row there should be read with this inversion in mind.

\subsection{The Glauber-inspired dynamics}
In Algorithm~\ref{algo:rheon} we provide the pseudocode for the generation of \eraclitus through \rheon dynamics as implemented in the released dataset.
At initialisation ($t=0$) every agent answers the question independently, with no peer context, yielding a non-interacting reference population.
The system is then evolved for $\bar N=10$ sweeps.
A single sweep performs $N_a$ elementary asynchronous updates:
each update (a) draws an agent $i$ uniformly at random, (b) shows it the \emph{current} responses of its neighbours $\mathcal N(i)$ together with its own previous answer, and (c) overwrites $r_i$ with the model's reply.

\begin{algorithm}[t]
\caption{The \rheon dynamics}
\label{algo:rheon}
\DontPrintSemicolon
\KwIn{prompt $p$; geometry $G$; size $N_a$; temperature $T$; sweeps $\bar N$}
\KwOut{responses $\{r_i(t)\}$ for $t=0,\dots,\bar N$}
$\mathcal N \leftarrow$ neighbourhoods induced by $(G,N_a)$\;
\tcp{Step 0: independent initialisation ($t=0$)}
\For{$i = 1 \dots N_a$}{
  $r_i \leftarrow \mathrm{LLM}\big(\mathrm{init}(p),\, T\big)$\;
}
\tcp{Steps $1,\dots,\bar N$: Glauber updates}
\For{$t = 1 \dots \bar N$}{
    \tcp{Asynchronous sub-steps}
    \For{$u = 1 \dots N_a$}{
        \tcp{(a) Select an agent}
        $i \sim \mathrm{Unif}\{1,\dots,N_a\}$\;
        \tcp{(b) Read live neighbours}
        $\mathrm{peers} \leftarrow \{\, r_j : j \in \mathcal N(i)\,\}$\;
        \tcp{(c) Update the answer}
        $r_i \leftarrow \mathrm{LLM}\big(\mathrm{ctx}(p, r_i, \mathrm{peers}),\, T\big)$\;
    }
}

\end{algorithm}

\subsection{The released artefact}
\eraclitusfull ships as one Parquet file per configuration and trial, named
\texttt{responses\_p\{p\}\_\allowbreak\{G\}\_\allowbreak L\{L\}\_\allowbreak T\{T\}\_\allowbreak gen\{r\}\allowbreak.parquet}, giving $432\times5=2{,}160$ files and ${\approx}39$\,GB in total.
Each file holds the $(1+\bar N)\,N_a$ rows of a single run --one row per generated response, the $t=0$ initialisation included-- under the schema of Table~\ref{tab:schema}.
Four conventions are worth stating explicitly, since each is easy to get wrong when reading the corpus:
the sweep index is stored as \texttt{t\_1}, not \texttt{t}, and runs $0,\dots,\bar N$;
the agent index \texttt{i} is $1$-based;
the \texttt{L} column is null for mean-field runs, which have no lattice side, so populations must be grouped by \texttt{N\_a} (the \texttt{L} field of a mean-field \emph{filename} carries $N_a$ instead);
and the trial label \texttt{gen\_index} takes the values $\{0,2,3,4,5\}$ over the five trials, the base run being $0$, so trials are most safely identified by filename.
Every row carries all four sentence-embedding columns, so the robustness check of Section~\ref{app:embedding-robustness} is reproducible without re-encoding the corpus.

\paragraph{Reconstructing the population state.}
Because a sweep performs $N_a$ updates drawn uniformly \emph{with replacement}, a stored row is an update \emph{event} rather than a snapshot of an agent:
within one sweep some agents are refreshed more than once and others not at all.
Measured over the released corpus, a sweep touches $63.8\pm3.9\%$ of the population, matching the $1-1/e\approx63.2\%$ expected of sampling with replacement.
Every observable in this paper is therefore computed on the \emph{carried-forward} state: the state of agent $i$ at sweep $t$ is its most recent stored response at or before $t$, so an agent left unselected during a sweep retains its previous answer instead of dropping out of the average.
Reading a sweep off the rows with \texttt{t\_1}$\,=t$ alone instead takes a ${\sim}64\%$ subsample biased towards the recently updated;
across configurations the two estimators of $\rho(\bar N)$ differ by $0.013$ on average --about three times more for $N_a\le64$ than for $N_a\ge216$-- but by as much as $0.66$ on an individual run, so the distinction matters most exactly where the main text reports bimodal outcomes.
Categories are then filtered as described above, $\rho(t)$ averaging over those agents whose carried-forward response is \emph{correct} or \emph{hallucinated}.

\begin{table}[t]
\centering
\footnotesize
\setlength{\tabcolsep}{4pt}
\begin{tabular}{@{}lp{0.58\columnwidth}@{}}
\toprule
Column & Contents \\
\midrule
\texttt{p}          & prompt identifier, $1$--$6$ \\
\texttt{geometry}   & \texttt{1D}, \texttt{2D}, \texttt{3D}, or \texttt{MF} \\
\texttt{L}          & lattice side; null for \texttt{MF} \\
\texttt{N\_a}       & population size \\
\texttt{T}          & sampling temperature \\
\texttt{t\_1}       & sweep index, $0$--$10$ \\
\texttt{i}          & agent index, $1$--$N_a$ \\
\texttt{gen\_index} & trial label, $\{0,2,3,4,5\}$ \\
\texttt{neighbors}  & indices of the agents whose responses were placed in this call's context; empty at $t=0$ \\
\texttt{ctx}        & the JSON user message given to the agent \\
\texttt{response}   & the generated text, exactly $100$ tokens \\
\texttt{category}   & judge verdict: $0$ correct, $1$ hallucinated, $2$ not-known, $-1$ judge-failure \\
\texttt{judge\_raw} & the judge's unparsed reply \\
\addlinespace
\textit{embeddings} & four columns, each named after its encoder: \MiniLMSix{} and \MiniLMTwelve{} ($384$ dimensions), \MpnetBase{} and \MpnetPara{} ($768$) \\
\bottomrule
\end{tabular}
\caption{Row schema of a released \eraclitusfull Parquet file. One row is one generated response, \ie one elementary update of the \rheon dynamics (Algorithm~\ref{algo:rheon}), together with its judge tag and its embeddings under all four encoders.}
\label{tab:schema}
\end{table}

\subsection{The LLM call budget}
Table~\ref{tab:LLM-calls} reports the LLM call budget of the campaign, organised by geometry. The budget is heavily asymmetric across the geometry ladder: the 3D lattice --the only geometry reaching $N_a=1000$-- accounts for $709{,}632$ of the $953{,}568$ calls of a single trial ($74.4\%$), while the 1D ring contributes just $2.3\%$.
Repeating the campaign for $R=5$ independent trials brings \eraclitusfull to $4{,}767{,}840$ generated responses; each is passed once more through the judge described above, with tagging details shown in Table~\ref{tab:judge-category-counts}.

\begin{table}[t]
\centering
\scriptsize
\begin{tabular}{lrrr}
\toprule
Geometry & $N_a$ & Calls & $\times 6$ prompts \\
\midrule
1D & $\{8,16,32\}$ & $3{,}696$ & $22{,}176$ \\
2D & $\{64,144,256\}$ & $30{,}624$ & $183{,}744$ \\
3D & $\{64,216,512,1000\}$ & $118{,}272$ & $709{,}632$ \\
Mean-field & $\{32,64\}$ & $6{,}336$ & $38{,}016$ \\
\midrule
Total (one trial) & & & $953{,}568$ \\
\textbf{Total ($R=5$ trials)} & & & $\mathbf{4{,}767{,}840}$ \\
\bottomrule
\end{tabular}
\caption{LLM call budget by geometry. Per each geometry it is specified the set of sizes $N_a$ used in the campaign, the number of calls for a single prompt summed over all sizes and temperatures, and the total multiplied by the six prompts. Totals are reported for one trial and for the full $R=5$ campaign.}
\label{tab:LLM-calls}
\end{table}

\section{Analysis of the LLM-as-a-judge tagging}
\label{sec:llm-tagging}

\begin{table}[t]
\centering
\footnotesize
\setlength{\tabcolsep}{4pt}
\begin{tabular}{lrrrrr}
\toprule
Geometry & Correct & Halluc. & Not-known & Judge-fail & Total \\
\midrule
1D   & 88{,}703 & 20{,}081 & 1{,}958 & 138 & \textbf{110{,}880} \\
2D   & 756{,}877 & 150{,}318 & 10{,}739 & 786 & \textbf{918{,}720} \\
3D   & 2{,}925{,}544 & 580{,}723 & 38{,}427 & 3{,}466 & \textbf{3{,}548{,}160} \\
MF   & 156{,}790 & 29{,}718 & 3{,}350 & 222 & \textbf{190{,}080} \\
\midrule
\textbf{Total} & \textbf{3{,}927{,}914} & \textbf{780{,}840} & \textbf{54{,}474} & \textbf{4{,}612} & \textbf{4{,}767{,}840} \\
\bottomrule
\end{tabular}

\caption{LLM-as-a-judge category counts by geometry, summed over the $432$ configurations ($6$ prompts, $12$ sizes, $6$ temperatures), all sweeps, and all \textbf{$R=5$ trials} of the released dataset, with per-geometry and per-category totals.
Hallucination shares quoted for this table in the text are taken over \emph{all} tagged rows, as in Table~\ref{tab:judge-category-counts-temp}, and so are not directly comparable with $\rho$ (see Table~\ref{tab:halluc-pct-prompt-temp}).}
\label{tab:judge-category-counts}
\end{table}

Table~\ref{tab:judge-category-counts} reports the judge-category counts by geometry, summed over every configuration and sweep of all five trials.
The corpus is largely \emph{correct} ($82.4\%$) with \emph{hallucinated} responses a substantial minority ($16.4\%$);
\emph{not-known} abstentions and \emph{judge-failures} are rare ($1.1\%$ and $0.1\%$ respectively).
The per-geometry totals track the asymmetry of the call budget (Table~\ref{tab:LLM-calls}) almost exactly, since the 3D lattice alone supplies $3{,}548{,}160$ of the $4{,}767{,}840$ tagged rows ($74.4\%$).
The hallucination share is close to the pooled average for 2D and 3D ($16.4\%$ each) and slightly lower for MF ($15.6\%$), but notably higher for 1D ($18.1\%$), the lowest-degree geometry.

\begin{table}[t]
\centering
\footnotesize
\setlength{\tabcolsep}{4pt}
\begin{tabular}{lrrrrr}
\toprule
$T$ & Correct & Halluc. & Not-known & Judge-fail & Halluc.\,\% \\
\midrule
0.1 & 691{,}107 & 93{,}661 & 9{,}403 & 469 & 11.8 \\
0.4 & 703{,}233 & 83{,}733 & 7{,}442 & 232 & 10.5 \\
0.7 & 643{,}959 & 142{,}643 & 7{,}637 & 401 & 18.0 \\
1 & 645{,}110 & 140{,}496 & 8{,}323 & 711 & 17.7 \\
1.3 & 635{,}996 & 145{,}443 & 11{,}535 & 1{,}666 & 18.3 \\
1.6 & 608{,}509 & 174{,}864 & 10{,}134 & 1{,}133 & 22.0 \\
\midrule
\textbf{Total} & \textbf{3{,}927{,}914} & \textbf{780{,}840} & \textbf{54{,}474} & \textbf{4{,}612} & \textbf{16.4} \\
\bottomrule
\end{tabular}
\caption{Judge-category counts by temperature, pooled over all geometries, prompts, sizes, sweeps, and all \textbf{$R=5$ trials}, with per-temperature and per-category totals.
\emph{Halluc.\,\%} is the share of \emph{all} tagged rows in the row, not of the adjudicated ones, and so is not directly comparable with $\rho$ (see Table~\ref{tab:halluc-pct-prompt-temp}).}

\label{tab:judge-category-counts-temp}
\end{table}

Table~\ref{tab:judge-category-counts-temp} reports the same judge-category counts pooled by sampling temperature instead of geometry.
The hallucination share climbs with $T$, from ${\sim}11\%$ at the coldest settings to $22.0\%$ at $T=1.6$.
\emph{Not-known} abstentions and \emph{judge-failures} remain rare throughout but are not flat:
judge-failures in particular grow with temperature (from $0.06\%$ at $T=0.1$ to a peak of $0.21\%$ at $T=1.3$), suggesting that hotter, more disordered generations are occasionally unparseable by the judge as well as more error-prone.

\begin{table}[t]
\centering
\footnotesize
\begin{tabular}{lrrrrrr}
\toprule
Prompt & T=0.1 & T=0.4 & T=0.7 & T=1 & T=1.3 & T=1.6 \\
\midrule
1 & 2.9 & 15.2 & 66.6 & 73.3 & 75.7 & 81.3 \\
2 & 3.7 & 4.2 & 10.0 & 10.4 & 10.3 & 9.0 \\
3 & 26.9 & 13.4 & 8.8 & 9.6 & 9.8 & 10.1 \\
4 & 0.0 & 0.0 & 0.0 & 0.0 & 0.5 & 1.9 \\
5 & 0.0 & 0.0 & 0.0 & 0.0 & 0.1 & 0.5 \\
6 & 39.2 & 31.2 & 23.7 & 15.1 & 17.8 & 33.2 \\
\bottomrule
\end{tabular}
\caption{Hallucination percentage by prompt and temperature, pooled over all configurations and all \textbf{$R=5$ trials} (the same full dataset as Tables~\ref{tab:judge-category-counts} and \ref{tab:judge-category-counts-temp}).
Here the percentage is taken over the \emph{adjudicated} responses only, $h/(c+h)$, matching the definition of $\rho$ in the main text.}
\label{tab:halluc-pct-prompt-temp}
\end{table}
Table~\ref{tab:halluc-pct-prompt-temp} shows that the pooled temperature-dependence of Table~\ref{tab:judge-category-counts-temp} masks substantially different per-prompt behaviour.
Prompt $1$ alone accounts for most of the aggregate trend, its tagged hallucination share rising almost monotonically from $2.9\%$ at $T=0.1$ to $81.3\%$ at $T=1.6$.
This is the outlier behaviour anticipated by the caveat on page~\pageref{par:prompt1-caveat}, and the reason the row carries no dynamical reading:
under the label inversion documented there the trend most plausibly runs the other way, hotter sampling producing more of the refutations that the judge misfiles as hallucinated.
We nonetheless do not relabel the corpus on the strength of a $200$-response audit, and draw no conclusion about the dynamics from this row.
Prompts $4$ and $5$ remain essentially hallucination-free ($\le 2\%$) across the whole range and only depart from zero at the two hottest temperatures, indicating a temperature floor below which these two questions are answered correctly with near-certainty rather than a gradual increase.
Prompts $3$ and $6$, by contrast, are \emph{non-monotonic} in the opposite sense to Prompt~1:
their hallucination percentage falls from $T=0.1$ to an intermediate temperature (around $T=0.7$--$1.0$) before climbing again at $T=1.6$.
Prompt~2 instead rises from a low value at the coldest settings to a mid-range plateau.
This heterogeneity shows that temperature alone does not determine the hallucination rate.
Its effect is nonetheless not symmetric once the mis-tagged Prompt~1 is set aside:
across Prompts~2--6 the range is dominated by repair, and the manufacture direction survives only as the mid-range plateau of Prompt~2 and as a minority of individual runs (about $5\%$ of the pooled runs in the main text).
What varies with $T$ is how much of the available error interaction removes, not which of the two directions it takes.

\begin{figure}[t]
    \centering
    \includegraphics[width=.5\linewidth]{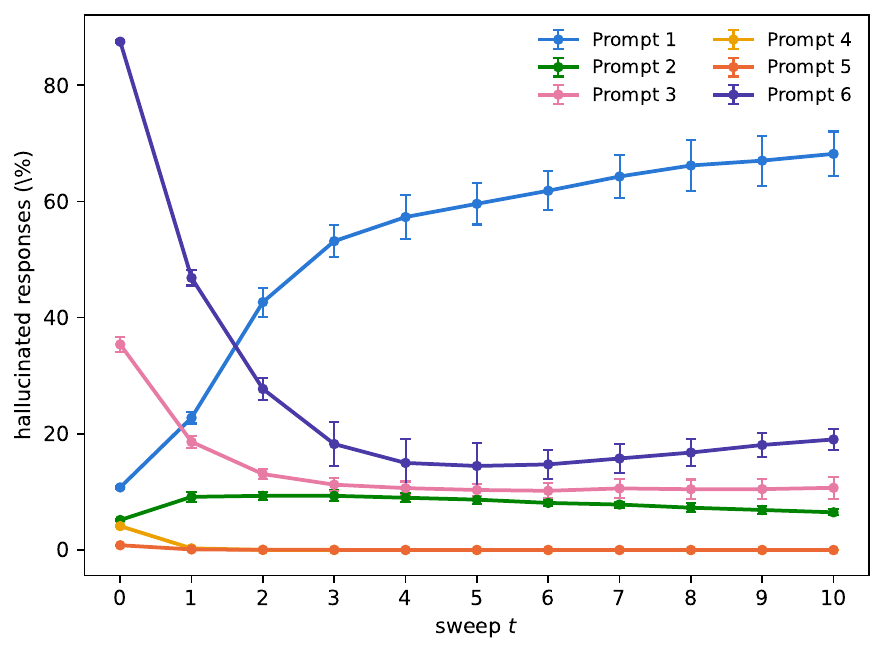}
\caption{Hallucination percentage for each prompt as a function of sweep $t$, pooled over all geometries, sizes, and temperatures. The marker is the mean over the $R=5$ trials; the error bars are $\pm1$ standard deviation across trials.}
\label{fig:halluc-pct-by-prompt}
\end{figure}

Figure~\ref{fig:halluc-pct-by-prompt} tracks the hallucination percentage of each prompt across the evolution sweep, pooled over geometry, size, and temperature, with the percentage formed within each trial before averaging so that the error bars measure trial-to-trial variability.
The six prompts fall into three regimes.
Prompt~1 rises steadily and almost monotonically, from $10.8\%$ at $t=0$ to $68.2\%$ at $t=10$; Prompts~3 and~6 show the opposite --and the largest-- effect: hallucination collapses within the first two sweeps ($35.4\%\to13.1\%$ and $87.5\%\to27.7\%$), bottoms out around $t=5$--$6$ ($10.2\%$ and $14.5\%$), and then drifts back up slightly (Prompt~6 to $19.0\%$ by $t=10$), so interaction acts as a fast repair mechanism that saturates, improving the factuality as shown in the main text.
Prompts~2,~4, and~5 barely move: Prompts~4 and~5 fall to $0\%$ within two sweeps and stay there, while Prompt~2 remains inside a $5$--$9\%$ band throughout.
The direction and magnitude of the interaction's effect on factuality is therefore prompt-specific rather than a uniform consequence of consensus-seeking. The trial-to-trial spread is small everywhere ($\sigma\le4.3$ percentage points, and $\le1.9$ outside Prompts~1 and~6), so these trajectories are stable across the five independent repetitions.
That stability is a property of the pooled curve and not of the individual configurations underneath it:
averaging over geometry, size, and temperature averages over the cold, densely-coupled runs whose outcome the main text reports as bimodal across trials.
The per-configuration dispersion is recovered in Figure~\ref{fig:mega-rho-evolution}, whose bands are formed within a single $(p,G,N_a,T)$ cell.

\section{Human validation protocol and results}\label{app:human-validation}

We validate the LLM-as-a-judge verdicts of Section~\ref{app:eraclitus} against human annotation on a stratified sample of the tagged corpus, drawn in two pools with a fixed random seed for extraction reproducibility:
\begin{enumerate}
    \item \textbf{Not-known oversample.} $200$ rows drawn uniformly at random from all rows tagged \emph{not-known} (category $2$), irrespective of configuration.
    This category is both the rarest ($1.1\%$ of the corpus, Table~\ref{tab:judge-category-counts}) and the hardest to adjudicate, so we oversample it deliberately.
    \item \textbf{Per-configuration sample.} $2$ rows drawn uniformly at random from \emph{each} $(p, G, N_a, T)$ configuration, so that every one of the $432$ configurations contributes to the audited sample regardless of its size or call volume.
\end{enumerate}
The union of the two pools yields $1{,}063$ responses (one configuration contributes a single row), each independently annotated by two humans (\emph{Ann.\,1} and \emph{Ann.\,2}) blind to the judge's verdict, using the same three-way scheme (\emph{correct} / \emph{hallucinated} / \emph{not-known}).
This design guarantees both breadth (every configuration is represented) and depth on the rarest category, at the cost of the sample no longer matching the corpus's raw category frequencies --an acceptable trade-off, since the goal is to estimate the judge's \emph{per-prompt} reliability, not to re-estimate the corpus-wide proportions already reported in Table~\ref{tab:judge-category-counts}.

\begin{table*}[htbp]
  \centering
  \small
  \begin{tabular}{lrccccccc}
    \toprule
    & & \multicolumn{3}{c}{Cohen's $\kappa$} & & \multicolumn{3}{c}{Raw agreement (\%)} \\
    \cmidrule(lr){3-5}\cmidrule(lr){7-9}
    Prompt & $N$ & LLM--A\,1 & LLM--A\,2 & A\,1--A\,2 & & LLM--A\,1 & LLM--A\,2 & A\,1--A\,2 \\
    \midrule
    1 & 200 & 0.025 & 0.366 & 0.015 & & 34.0 & 58.5 & 33.5 \\
    2 & 146 & 1.000 & 0.880 & 0.880 & & 100.0 & 97.9 & 97.9 \\
    3 & 267 & 1.000 & 0.895 & 0.895 & & 100.0 & 94.0 & 94.0 \\
    4 & 143 & -- & -- & -- & & 100.0 & 100.0 & 100.0 \\
    5 & 144 & -- & -- & -- & & 100.0 & 100.0 & 100.0 \\
    6 & 163 & 1.000 & 0.897 & 0.897 & & 100.0 & 93.9 & 93.9 \\
    \midrule
    All & 1063 & 0.760 & 0.791 & 0.707 & & 87.6 & 89.5 & 84.8 \\
    \bottomrule
  \end{tabular}
  \caption{Inter-rater agreement between the LLM judge and the two independent human annotators (\emph{A\,1}, \emph{A\,2}), per prompt and pooled, as Cohen's $\kappa$ (left) and as raw percentage agreement (right). Dashes mark prompts where all three raters assigned the same label to every response, leaving $\kappa$ undefined but raw agreement well defined at $100\%$.}
  \label{tab:agreement}
\end{table*}

Table~\ref{tab:agreement} reports Cohen's $\kappa$ between the judge and each human annotator, and between the two annotators, per prompt and pooled.
Pooled over the whole sample the judge and the humans agree on $87.6\%$ of responses, with $\kappa=0.760$ and $0.791$ against the two annotators --both above the $\kappa=0.707$ the annotators reach with \emph{each other}, so the judge is about as consistent with a human as two humans are with one another.
On Prompts~2--6 the judge is essentially perfect: raw agreement with Ann.\,1 is $100\%$ on every one of them, with $\kappa=1.0$ on Prompts~2, 3, and~6 and $\kappa$ undefined on Prompts~4 and~5 (there every rater tagged every sampled response \emph{correct}, leaving no variance to correlate).
Averaging those per-prompt raw agreements over both annotators gives $98.58\%$, the shortfall being carried entirely by Ann.\,2 on Prompts~2, 3, and~6.
Prompt~1 is the lone exception: agreement with Ann.\,1 collapses to $34\%$ and $\kappa$ to $0.025$ --the judge is statistically \emph{uncorrelated} with the human labels there.
This is the quantitative form of the false-premise failure discussed on page~\pageref{par:prompt1-caveat}: because the judge cannot recognise that no cube is the sum of two positive cubes, it tags correct refutations of Prompt~1 as hallucinated, disagreeing with the humans almost at chance.
It is exactly this measurement that licenses the main text's decision to exclude Prompt~1 from every pooled analysis while retaining Prompts~2--6, on which the automatic tags are validated as human-level.

\paragraph{The Prompt-1 disagreement is an inversion, not noise.}
The confusion matrix separates the two directions of that disagreement.
Of the $200$ audited Prompt-1 responses, the judge and Ann.\,1 agree on all $58$ tagged \emph{not-known}, but on the adjudicated axis they agree on only $10$ of $142$:
the judge calls $82$ responses \emph{correct} that Ann.\,1 calls hallucinated, and $50$ \emph{hallucinated} that Ann.\,1 calls correct.
Reading the responses in those two cells settles which rater is right.
All $82$ assert the taxicab number $1729$, which is a sum of two positive cubes in two ways but is not itself a cube;
the $50$ are refutations, stating that no such cube exists.
Ann.\,1 is thus correct on every adjudicable case and the judge wrong on all of them, and exchanging the judge's \emph{correct} and \emph{hallucinated} labels on this prompt alone raises its agreement with Ann.\,1 from $34.0\%$ to $95.0\%$, and $\kappa$ from $0.025$ to $0.923$.
The same exchange \emph{lowers} agreement with Ann.\,2 to $32.5\%$ ($\kappa=-0.002$), which is what identifies Ann.\,1 as the reference annotator on this prompt:
Ann.\,2 accepted the false premise alongside the judge, so the low Ann.\,1--Ann.\,2 agreement of Table~\ref{tab:agreement} records that error rather than genuine ambiguity in the question.
We report the inversion rather than act on it:
a relabelling inferred from $200$ audited responses should not be propagated to the $794{,}640$ rows the prompt contributes, so Prompt~1 is excluded rather than repaired.

\section{Order parameters}\label{app:order-parameter}
In this section we first give a graphical reading of the order parameters introduced in the main text --the magnetisation $m(t)$, the cohesion $\bar c(t)$, and the hallucination density $\rho(t)$-- and then extend the main text's trajectory, phase-space, and temperature figures to all five analysed prompts.

\subsection{A geometric interpretation of the order parameters}
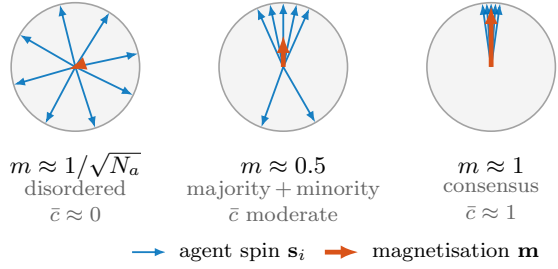
\begin{figure}[t]
    \centering
    \input{figs/order-schematic}
    \caption{
        Schematic of the semantic order parameters on the unit sphere, shown in 2D projection.
        Thin arrows are the agent spins $\mathbf s_i$;
        the thick arrow is the semantic magnetisation $\mathbf m$, whose length is $m$.
        \emph{Left:} a disordered population, spins spread over the sphere, $m$ near its $1/\sqrt{N_a}$ floor and cohesion $\bar c\approx0$.
        \emph{Centre:} partial order, $m\approx0.5$, a majority direction with a dissenting minority.
        \emph{Right:} consensus, spins aligned, $m\approx1$ and $\bar c\approx1$.
    }
    \label{fig:order-schematic}
\end{figure}

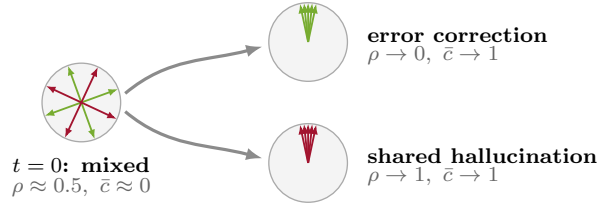
\begin{figure}[t]
    \centering
    \input{figs/factual-outcomes}
    \caption{
        The two consensus outcomes.
        A population that starts mixed and disordered (left, $\rho\approx0.5$, $\bar c\approx0$) can settle, at high cohesion, into either collective error correction (top, $\rho\to0$) or a shared hallucination (bottom, $\rho\to1$).
        Arrows are agent spins, coloured by factual state (green: correct; red: hallucinated);
        which branch a configuration takes is not fixed by its initial accuracy (see the main-text results).
    }
    \label{fig:factual-outcomes}
\end{figure}

Figure~\ref{fig:order-schematic} gives a geometric reading of the two order parameters defined in the main text.
Each agent's response is embedded and normalised to a unit vector $\mathbf s_i$, so a population is a cloud of points on the unit sphere --drawn here in 2D projection-- and the semantic magnetisation $\mathbf m$ is their mean resultant, whose length $m$ measures how far that cloud departs from isotropy.
The left panel depicts a disordered reference: the spins spread over the sphere, the resultant very nearly cancels, and $m$ settles at its finite-size floor $1/\sqrt{N_a}$ rather than at zero --precisely the offset that the cohesion $\bar c=(N_a m^2-1)/(N_a-1)$ subtracts.
The centre panel shows partial order, a majority direction with a dissenting minority; since $\bar c\approx m^2$ at large $N_a$, a population at $m\approx0.5$ is only moderately cohesive ($\bar c\approx0.25$), so the two parameters do not advance on the same scale.
The right panel represent consensus: the spins are nearly parallel, the resultant is close to unit length, and $m$ and $\bar c$ both approach $1$.
What the schematic deliberately does not encode is the \emph{content} of the direction the population settles on --any of the three panels is compatible with any level of factual accuracy-- which is the distinction Figure~\ref{fig:factual-outcomes} makes explicit.

Figure~\ref{fig:factual-outcomes} adds the factual axis that Figure~\ref{fig:order-schematic} omits, colouring each spin by the judge's verdict on the response it encodes.
A population that begins mixed and disordered (left) can end in either of two states, and the essential point is that the two right-hand panels are \emph{geometrically identical}: both have their spins nearly parallel, both sit at $m\approx1$ and $\bar c\approx1$, and nothing in the order parameters distinguishes them.
Only the colouring does: $\rho\to0$ above, collective error correction, against $\rho\to1$ below, a shared hallucination in which every agent gives the same confident wrong answer.
Consensus is therefore degenerate on the semantic axes alone, which is why $\rho$ is tracked as an independent observable rather than inferred from cohesion, and why the correlations between $\bar c$ and $1-\rho$ reported in the main text are, a priori, positive but weak.

\subsection{Evolution of the order parameters across the prompts}
\begin{figure*}[p]
    \centering
    \includegraphics[width=\textwidth]{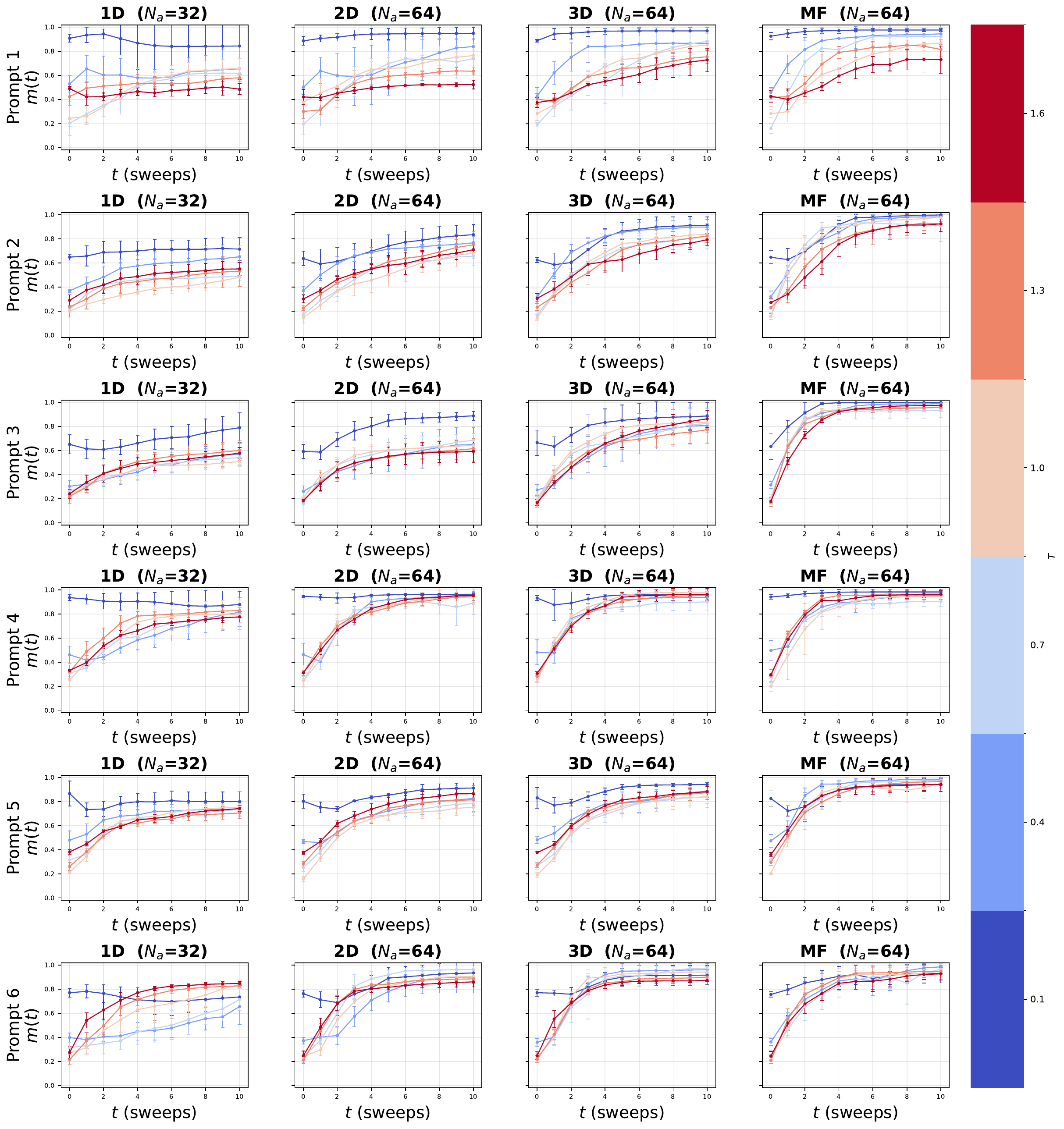}
    \caption{Semantic magnetisation $m(t)$ for all five analysed prompts (Prompts~2--6, rows) across all four geometries (columns), one curve per temperature.}
    \label{fig:mega-m-evolution}
\end{figure*}

Figure~\ref{fig:mega-m-evolution} extends the main-text magnetisation evolution (shown there only for Prompt 3) to all five analysed prompts simultaneously, arranged as a $5 \times 4$ grid with prompts as rows and geometries as columns (Prompt~1 is excluded throughout, as everywhere else in the analysis).
Within each panel, one curve per temperature traces the mean semantic magnetisation $m(t)$ over the $t=0,\dots,10$ sweeps, with the shaded band showing the standard deviation $\pm$ std across the independent runs available for that configuration.
Reading across a row shows how a single prompt's ordering dynamics depend on connectivity, holding the question fixed;
reading down a column shows how a single geometry's convergence behaviour varies across prompts, holding the interaction structure fixed.

As in the main text, magnetisation rises from its $t=0$ baseline toward a plateau within the first few sweeps in every panel, with the rate and asymptotic level set by geometry: higher connectivity orders faster and further, along the ladder from 1D through mean-field.
The curves also separate by temperature, but that separation should not be read as colder populations ordering more completely.
As the main text shows, the cold runs \emph{start} far higher --near-greedy sampling makes independent $t=0$ responses nearly identical before any interaction has occurred-- so the cold curves sit above the others throughout without interaction having done more work for them.
On this observable the temperature ordering is largely inherited from $t=0$ rather than produced by the dynamics, which is why the main text poses its temperature question on the factual axis, where the initial level can be differenced out.
The grid format makes it possible to check, prompt by prompt, whether the geometry pattern --established on Prompt~3 alone in the main text-- generalises across the analysed prompts, and to spot any prompt whose ordering trajectory departs from it.

\begin{figure*}[p]
    \centering
    \includegraphics[width=\textwidth]{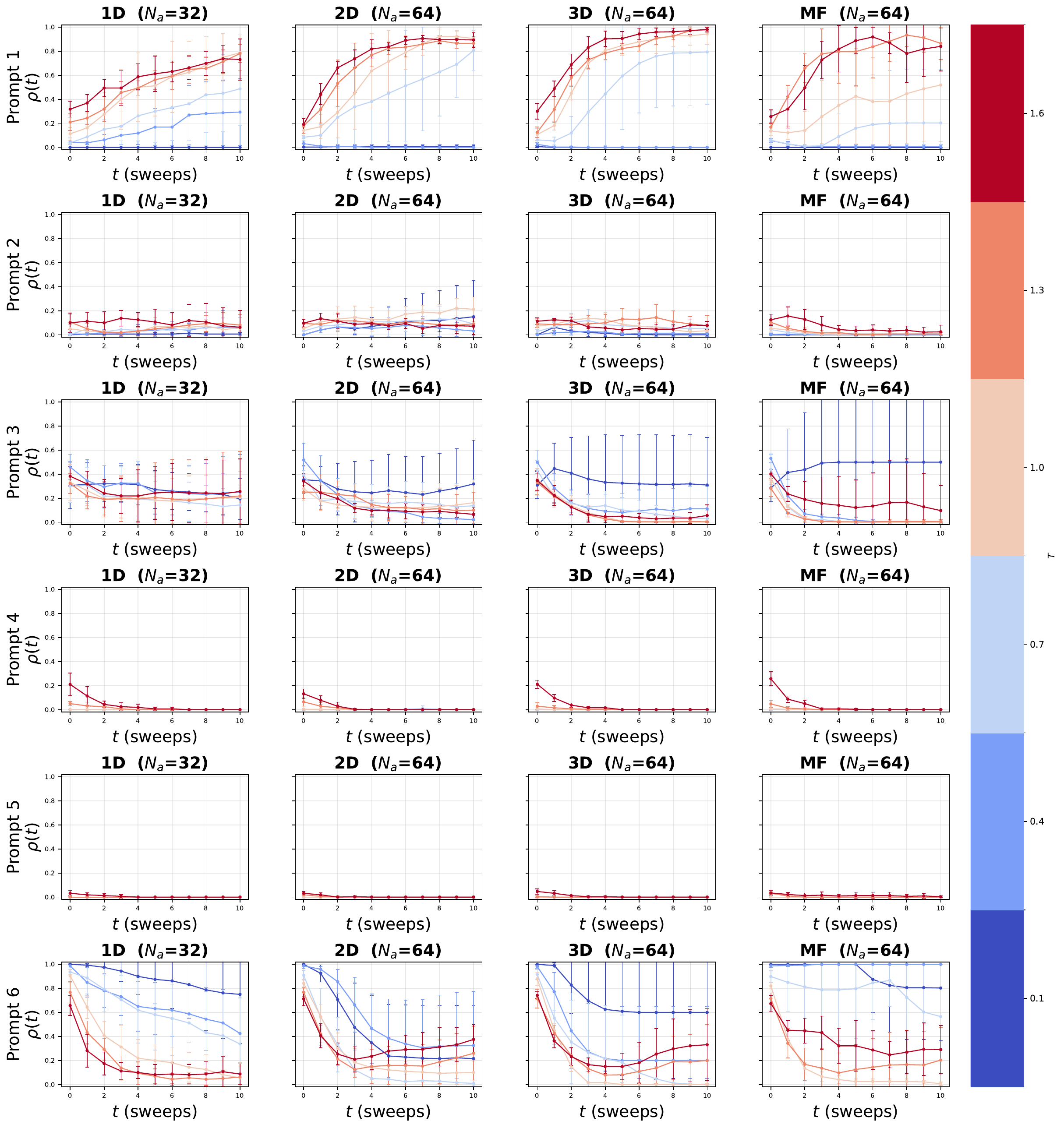}
    \caption{Hallucination density $\rho(t)$ for all five analysed prompts (Prompts~2--6, rows) across all four geometries (columns), one curve per temperature.}
    \label{fig:mega-rho-evolution}
\end{figure*}

Figure~\ref{fig:mega-rho-evolution} shows a grid for the hallucination density $\rho(t)$, using the same $5 \times 4$ (prompt $\times$ geometry) layout, temperature colour-coding, and mean $\pm$ std band convention as per Figure~\ref{fig:mega-m-evolution}.
Where the magnetisation grid shows \emph{how strongly} the population orders, this grid shows \emph{what it orders onto}:
whether the consensus the agents converge toward is factually correct or hallucinated.

Comparing the two mega-grids panel by panel is the main use of these figures:
a prompt/geometry combination can show strong magnetisation growth in Figure~\ref{fig:mega-m-evolution} while its $\rho(t)$ trace in Figure~\ref{fig:mega-rho-evolution} rises rather than falls, indicating that the population is converging confidently onto a shared \emph{wrong} answer rather than a correct one.
This is the grid-level counterpart to the per-prompt hallucination trajectories already reported in Table~\ref{tab:halluc-pct-prompt-temp}/Figure~\ref{fig:halluc-pct-by-prompt} (pooled over geometry and temperature), now disaggregated so the role of connectivity and temperature within each prompt is visible directly.

\begin{figure*}[p]
    \centering
    \includegraphics[width=0.95\textwidth,height=0.88\textheight,keepaspectratio]{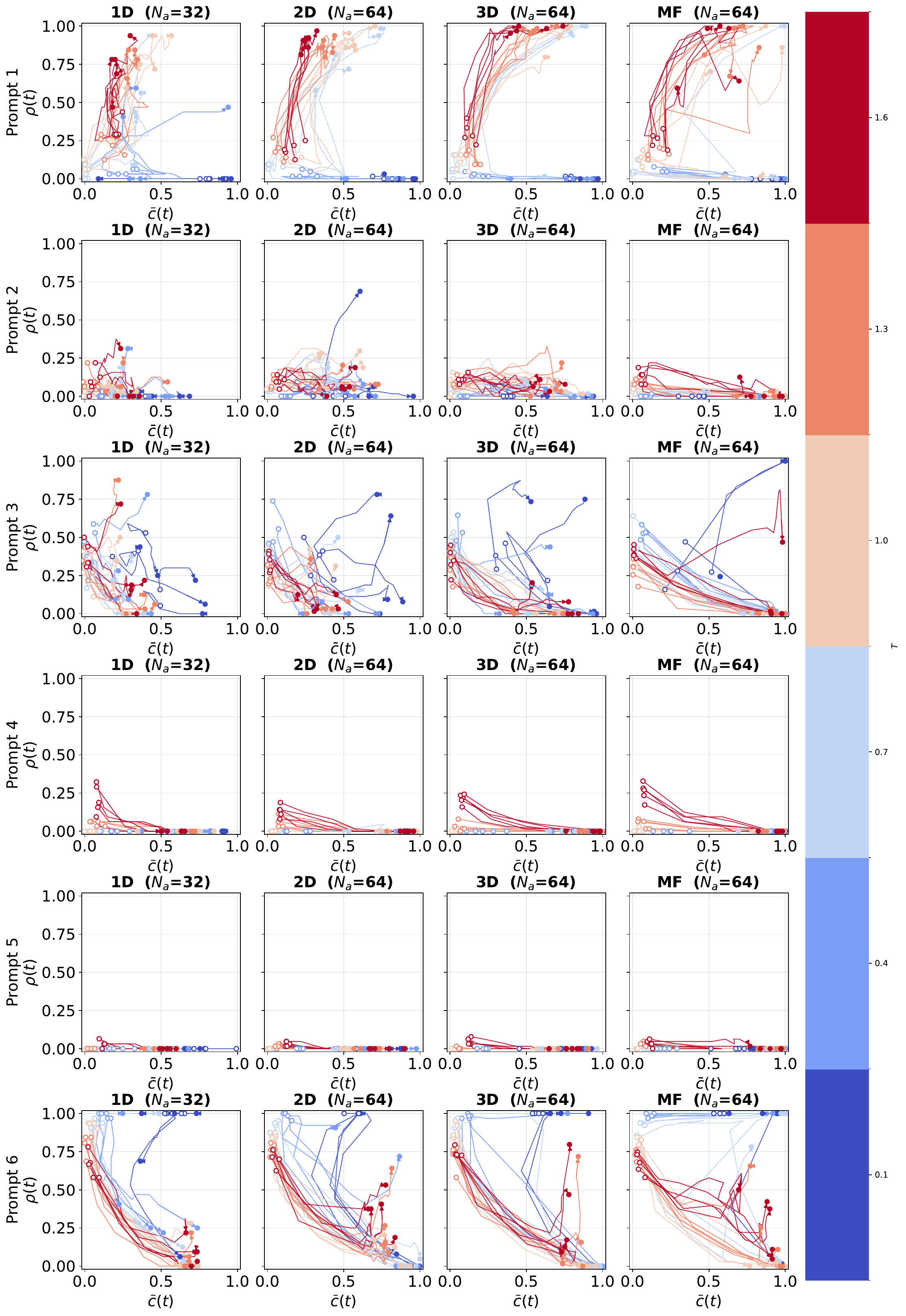}
    \caption{Phase-space trajectories $(\bar c(t), \rho(t))$ for all five analysed prompts (Prompts~2--6, rows) across all four geometries (columns), one curve per temperature.}
    \label{fig:mega-phase-geom}
\end{figure*}

Figure~\ref{fig:mega-phase-geom} extends the main-text phase-space trajectories (there being Prompt~3 only) to all five analysed prompts, again as a $5\times 4$ (prompt $\times$ geometry) grid.
Each panel plots the trajectory of $(\bar c(t), \rho(t))$ --semantic cohesion on the horizontal axis, hallucination density on the vertical-- as the population evolves from $t=0$ (hollow circle) to $t=10$ (filled circle, with an arrowhead on the final segment), one trajectory per run per temperature, coloured by $T$ as in the other plots.

This joint view separates two questions that the two evolution grids above address separately:
how much the population agrees ($\bar c$) and whether that agreement is correct ($\rho$).
A trajectory moving right and down indicates cohesion increasing while hallucination falls --healthy consensus-- whereas a trajectory moving right and up indicates the population compacting onto a shared error.
Because every panel shares the same axes and endpoint-marker convention, the grid supports a direct visual comparison of which prompt/geometry combinations land in which quadrant of the $(\bar c, \rho)$ plane by the end of the run.
It is also where the main text's central qualitative claim can be checked prompt by prompt:
the cold, densely-coupled trajectories that peel away toward the top-right corner --high cohesion at high hallucination-- are the concrete instances of a population being most unanimous exactly where it is least reliable.

\begin{figure*}[p]
    \centering
    \includegraphics[width=\textwidth,height=0.88\textheight,keepaspectratio]{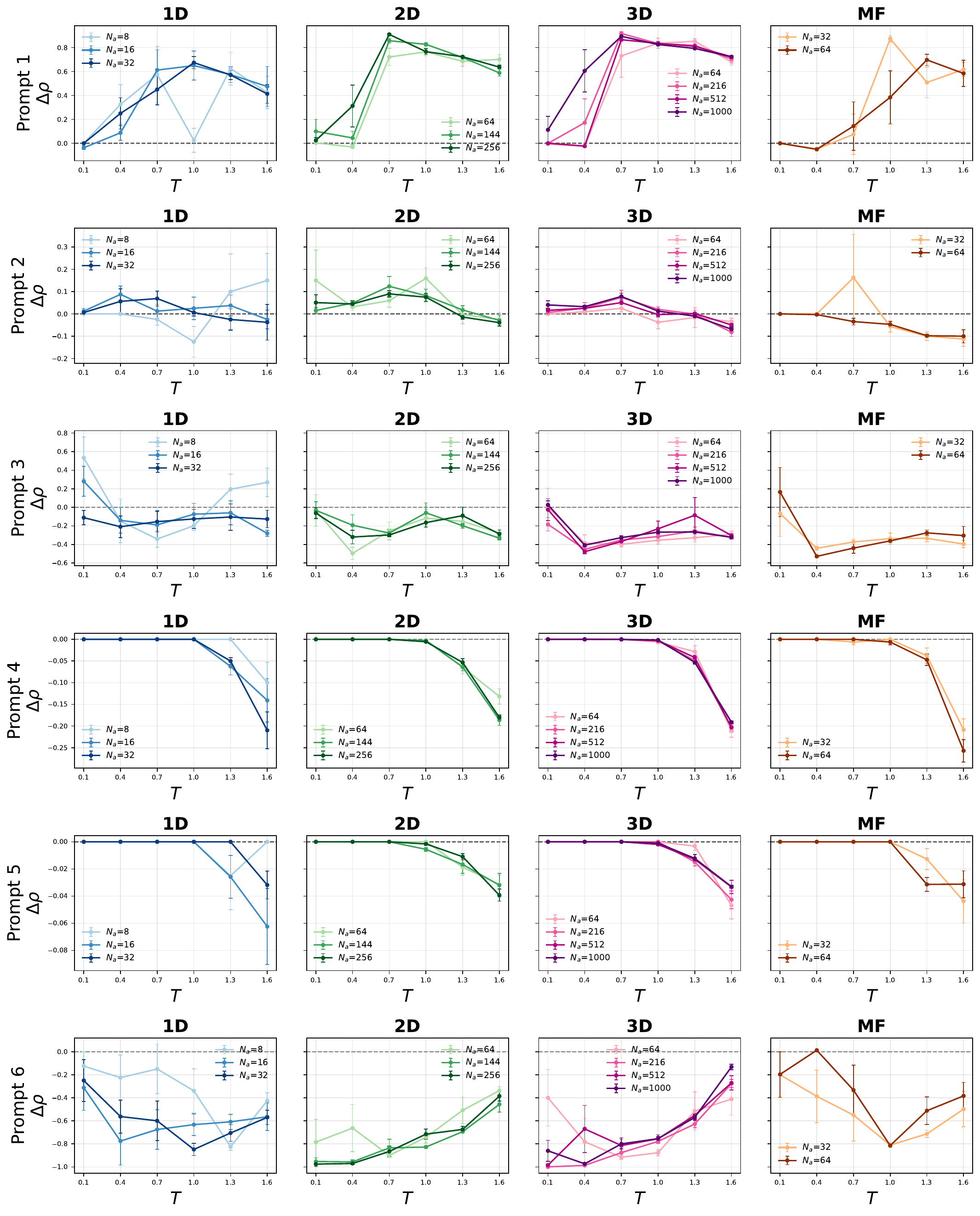}
    \caption{Interaction-induced change in hallucination, $\Delta\rho=\rho(t{=}10)-\rho(t{=}0)$, against sampling temperature $T$ for all five analysed prompts (Prompts~2--6, rows) across all four geometries (columns).
    One curve per system size, on that geometry's colour ramp from light (small) to dark (large);
    error bars are the standard error over the $R=5$ trials, and the dashed line marks the $\Delta\rho=0$ expected of a symmetric voter model.
    The vertical scale is shared within a row but not across rows, since $\Delta\rho$ spans a very different range from prompt to prompt.}
    \label{fig:mega-temperature-geom}
\end{figure*}

Figure~\ref{fig:mega-temperature-geom} extends the main text's temperature figure (drawn there for Prompt~6 alone) to all five analysed prompts, in the same $5\times4$ (prompt $\times$ geometry) layout as the grids above.
Unlike the three preceding plots, though, colour here encodes system size rather than temperature, with $T$ being the horizontal axis;
each panel therefore carries its geometry's own size ladder ($N_a\in\{8,16,32\}$ for 1D, $\{64,144,256\}$ for 2D, $\{64,216,512,1000\}$ for 3D, $\{32,64\}$ for MF).

\section{Embedding-robustness results}\label{app:embedding-robustness}

Every observable in the main text is computed from a single sentence encoder, \MiniLMSix{}.
Because the semantic spin --and hence every order parameter-- is a deterministic function of that embedding, we verify that the consensus phenomenology is a property of the population dynamics rather than an artefact of one particular encoder.
We re-embed all ${\sim}4.7$M responses under three further encoders --\MiniLMTwelve{}, \MpnetBase{}, and \MpnetPara{}-- and re-run the entire pipeline (per-prompt whitening~\citep{whitening2021}, unit-spin projection, and the order-parameter estimators) unchanged, once per encoder.
Two checks summarise the outcome, which, taken together, show that the observables and the conclusions drawn from them are robust to the choice of sentence encoder: switching encoder rescales the order parameters slightly but leaves the geometry ladder and the per-configuration ranking essentially unchanged. In the tables below, the encoders are abbreviated \texttt{MiniLM-6}/\texttt{-12} and \texttt{MPNet-b}/\texttt{-p}.

\subsection{The geometry ladder is preserved}
As shown in Table~\ref{tab:robust-ladder},
final semantic cohesion $\bar c_{\mathrm{final}}$ increases monotonically along the geometry ladder 1D $<$ 2D $<$ 3D $<$ MF under all four encoders, with mean-field reaching $\bar c_{\mathrm{final}}\approx0.92$ in every case.
The absolute levels shift slightly between encoders --\MpnetPara{} runs a few points higher on the low-degree lattices, for instance-- but the ordering that carries the paper's central structural claim is identical across encoders.

\subsection{The per-configuration ordering is preserved}
Beyond the geometry averages, Table~\ref{tab:robust-spearman} shows instead that the final cohesion of each individual configuration is stable across encoders: the Spearman rank correlation of $\bar c_{\mathrm{final}}$ between any two encoders, computed over all $(\text{config},\text{trial})$ pairs, lies between $0.899$ and $0.962$.
The encoders therefore agree not only on the geometry ladder but on the fine-grained ranking of which configurations cohere most strongly.

\begin{table}[t]
\centering
\small
\begin{tabular}{lrrrr}
\toprule
Geometry & MiniLM-6 & MiniLM-12 & MPNet-b & MPNet-p \\
\midrule
1D & 0.621 & 0.613 & 0.592 & 0.667 \\
2D & 0.634 & 0.635 & 0.609 & 0.710 \\
3D & 0.731 & 0.723 & 0.704 & 0.782 \\
MF & 0.926 & 0.920 & 0.916 & 0.919 \\
\bottomrule
\end{tabular}
\caption{Final semantic cohesion $\bar c_{\mathrm{final}}$ by geometry under each of the four sentence encoders, averaged over prompts (2--6), sizes, temperatures and trials. The geometry ladder 1D $<$ 2D $<$ 3D $<$ MF holds under every encoder.}
\label{tab:robust-ladder}
\end{table}

\begin{table}[t]
\centering
\small
\begin{tabular}{lrrrr}
\toprule
 & MiniLM-6 & MiniLM-12 & MPNet-b & MPNet-p \\
\midrule
MiniLM-6 & 1.000 & 0.962 & 0.934 & 0.916 \\
MiniLM-12 & 0.962 & 1.000 & 0.947 & 0.899 \\
MPNet-b & 0.934 & 0.947 & 1.000 & 0.923 \\
MPNet-p & 0.916 & 0.899 & 0.923 & 1.000 \\
\bottomrule
\end{tabular}
\caption{Spearman rank correlation of the per-configuration final cohesion $\bar c_{\mathrm{final}}$ between the four encoders, over all $(\text{config},\text{trial})$ pairs. Values $\ge0.89$ throughout: the ordering of configurations by cohesion is essentially encoder-invariant.}
\label{tab:robust-spearman}
\end{table}

\end{document}

%% file: figs/geometry-ladder.tex
\begin{tikzpicture}[
    refnode/.style={circle, draw=black, fill=redM!80, minimum size=2.5mm, inner sep=0pt, line width=0.4pt},
    nbrnode/.style={circle, draw=black, fill=blueM!65, minimum size=2.5mm, inner sep=0pt, line width=0.4pt},
    fadenode/.style={circle, draw=greyYed!40, fill=blueM!16, minimum size=2.5mm, inner sep=0pt, line width=0.3pt},
    refnode3/.style={circle, draw=black, fill=redM!80, minimum size=1.9mm, inner sep=0pt, line width=0.3pt},
    nbrnode3/.style={circle, draw=black, fill=blueM!65, minimum size=1.9mm, inner sep=0pt, line width=0.3pt},
    fadenode3/.style={circle, draw=greyYed!40, fill=blueM!16, minimum size=1.9mm, inner sep=0pt, line width=0.25pt},
    gedge/.style={greyYed!70, line width=0.7pt},
    gedge3/.style={greyYed!50, line width=0.4pt},
    gmf/.style={greyYed!40, line width=0.35pt},
    hledge/.style={blueM!70, line width=1.0pt},
    hledge3/.style={blueM!70, line width=0.7pt},
    gwrap/.style={greyYed!85, line width=0.3pt, -{Latex[length=0.6mm]}},
    plab/.style={font=\scriptsize\bfseries, text=black},
    slab/.style={font=\scriptsize, text=greyYed},
    nlab/.style={font=\scriptsize, text=greyYed},
    dimlab/.style={font=\scriptsize\itshape, text=greyYed}
]
    \def\rr{0.725}
    \def\gs{0.44}
    \def\xA{0}
    \def\xB{1.9}
    \def\xC{4.35}
    \def\xD{6.65}

    \begin{scope}[shift={(\xA,0)}]
        \foreach \i in {1,...,6}{ \coordinate (r\i) at ({(\i-1)*60}:\rr-0.07); }
        \foreach \i [evaluate=\i as \n using {int(mod(\i,6)+1)}] in {1,...,6}{
            \draw[gedge] (r\i) -- (r\n);
        }
        \draw[hledge] (r1)--(r2);  \draw[hledge] (r1)--(r6);
        \foreach \i in {1,...,6}{ \node[fadenode] at (r\i) {}; }
        \node[nbrnode] at (r2) {}; \node[nbrnode] at (r6) {};
        \node[refnode] at (r1) {};
        \node[nlab] at (0,\rr+0.34) {$\{8,16,32\}$};
        \node[plab] at (0,-\rr-0.30) {1D ring};
        \node[slab] at (0,-\rr-0.58) {$k=2$};
    \end{scope}

    \begin{scope}[shift={(\xB,0)}]
        \foreach \c in {0,1,2}{ \foreach \r in {0,1,2}{
            \coordinate (t\c\r) at (\c*\gs-\gs, \r*\gs-\gs);
        }}
        \foreach \r in {0,1,2}{ \draw[gedge] (t0\r)--(t1\r)--(t2\r); }
        \foreach \c in {0,1,2}{ \draw[gedge] (t\c0)--(t\c1)--(t\c2); }
        \foreach \r in {0,1,2}{
            \draw[gwrap] (t0\r) -- ++(-0.26,0);
            \draw[gwrap] (t2\r) -- ++(0.26,0);
        }
        \foreach \c in {0,1,2}{
            \draw[gwrap] (t\c0) -- ++(0,-0.26);
            \draw[gwrap] (t\c2) -- ++(0,0.26);
        }
        \draw[hledge] (t11)--(t01); \draw[hledge] (t11)--(t21);
        \draw[hledge] (t11)--(t10); \draw[hledge] (t11)--(t12);
        \foreach \c in {0,1,2}{ \foreach \r in {0,1,2}{ \node[fadenode] at (t\c\r) {}; }}
        \node[nbrnode] at (t01) {}; \node[nbrnode] at (t21) {};
        \node[nbrnode] at (t10) {}; \node[nbrnode] at (t12) {};
        \node[refnode] at (t11) {};
        \node[nlab] at (0,\rr+0.34) {$\{64,144,256\}$};
        \node[plab] at (0,-\rr-0.30) {2D torus};
        \node[slab] at (0,-\rr-0.58) {$k=4$};
    \end{scope}

    \begin{scope}[shift={(\xC,0)}]
      \def\p{0.46}
      \def\qx{0.66}
      \def\qy{0.30}
      \begin{scope}[shift={(-\p-\qx,-\p-\qy)}]
        \foreach \k in {0,1,2}{ \foreach \i in {0,1,2}{ \foreach \j in {0,1,2}{
            \coordinate (n\i\j\k) at (\i*\p+\k*\qx, \j*\p+\k*\qy);
        }}}
        \foreach \k in {0,1,2}{ \foreach \j in {0,1,2}{ \draw[gedge3] (n0\j\k)--(n1\j\k)--(n2\j\k); }}
        \foreach \k in {0,1,2}{ \foreach \i in {0,1,2}{ \draw[gedge3] (n\i0\k)--(n\i1\k)--(n\i2\k); }}
        \foreach \i in {0,1,2}{ \foreach \j in {0,1,2}{ \draw[gedge3] (n\i\j0)--(n\i\j1)--(n\i\j2); }}
        \draw[gwrap] (n211) -- ++(0.24,0);
        \draw[gwrap] (n011) -- ++(-0.24,0);
        \draw[gwrap] (n121) -- ++(0,0.24);
        \draw[gwrap] (n101) -- ++(0,-0.24);
        \draw[gwrap] (n112) -- ++(0.22,0.10);
        \draw[gwrap] (n110) -- ++(-0.22,-0.10);
        \draw[hledge3] (n111)--(n011); \draw[hledge3] (n111)--(n211);
        \draw[hledge3] (n111)--(n101); \draw[hledge3] (n111)--(n121);
        \draw[hledge3] (n111)--(n110); \draw[hledge3] (n111)--(n112);
        \foreach \k in {0,1,2}{ \foreach \i in {0,1,2}{ \foreach \j in {0,1,2}{
            \node[fadenode3] at (n\i\j\k) {};
        }}}
        \foreach \nb in {n011,n211,n101,n121,n110,n112}{ \node[nbrnode3] at (\nb) {}; }
        \node[refnode3] at (n111) {};
      \end{scope}
        \node[nlab] at (0,\rr+0.34) {$\{64,216,512,1000\}$};
        \node[plab] at (0,-\rr-0.30) {3D torus};
        \node[slab] at (0,-\rr-0.58) {$k=6$};
    \end{scope}

    \begin{scope}[shift={(\xD,0)}]
        \foreach \i in {1,...,6}{ \coordinate (m\i) at ({(\i-1)*60}:\rr-.05); }
        \foreach \i in {1,...,6}{ \foreach \j in {1,...,6}{ \draw[gmf] (m\i)--(m\j); }}
        \foreach \i in {2,...,6}{ \draw[hledge] (m1)--(m\i); }
        \foreach \i in {2,...,6}{ \node[nbrnode] at (m\i) {}; }
        \node[refnode] at (m1) {};
        \node[nlab] at (0,\rr+0.34) {$\{32,64\}$};
        \node[plab] at (0,-\rr-0.30) {mean-field};
        \node[slab] at (0,-\rr-0.58) {$k=N_a-1$};
    \end{scope}

    \draw[-{Latex[length=2mm]}, greyYed!80, line width=0.8pt]
        (\xA-\rr-0.15,-\rr-0.90) -- (\xD+\rr+0.15,-\rr-0.90);
    \node[dimlab] at ({0.5*\xD},-\rr-1.10) {increasing effective dimension};
\end{tikzpicture}

%% file: figs/order-schematic.tex
\begin{tikzpicture}[
    spin/.style={-{Latex[length=1.5mm]}, blueM, line width=0.7pt, opacity=0.9},
    resultant/.style={-{Latex[length=2.4mm]}, orangeM, line width=1.8pt},
    disc/.style={draw=greyYed!60, fill=greyYed!7, line width=0.6pt},
    plab/.style={font=\small\bfseries, text=black},
    slab/.style={font=\footnotesize, text=greyYed, align=center},
    key/.style={font=\footnotesize, text=black, anchor=west}
]
    \def\R{0.85}
    \def\dx{2.75}
    \begin{scope}[shift={(0,0)}]
        \draw[disc] (0,0) circle (\R);
        \foreach \a in {12,58,104,150,196,242,288,334}{
            \draw[spin] (0,0) -- (\a:\R);
        }
        \draw[resultant] (0,0) -- (200:0.12*\R);
        \node[plab] at (0,-\R-0.45) {$m \approx 1/\sqrt{N_a}$};
        \node[slab] at (0,-\R-0.92) {disordered\\$\bar c \approx 0$};
    \end{scope}
    \begin{scope}[shift={(\dx,0)}]
        \draw[disc] (0,0) circle (\R);
        \foreach \a in {66,78,90,102,114}{
            \draw[spin] (0,0) -- (\a:\R);
        }
        \foreach \a in {250,300}{
            \draw[spin] (0,0) -- (\a:\R);
        }
        \draw[resultant] (0,0) -- (90:0.5*\R);
        \node[plab] at (0,-\R-0.45) {$m \approx 0.5$};
        \node[slab] at (0,-\R-0.92) {majority\,+\,minority\\$\bar c$ moderate};
    \end{scope}
    \begin{scope}[shift={(2*\dx,0)}]
        \draw[disc] (0,0) circle (\R);
        \foreach \a in {82,86,90,94,98}{
            \draw[spin] (0,0) -- (\a:\R);
        }
        \draw[resultant] (0,0) -- (90:0.95*\R);
        \node[plab] at (0,-\R-0.45) {$m \approx 1$};
        \node[slab] at (0,-\R-0.92) {consensus\\$\bar c \approx 1$};
    \end{scope}
    \begin{scope}[shift={(\dx,-\R-1.6)}]
        \draw[spin] (-2.0,0) -- (-1.55,0);
        \node[key] at (-1.5,0) {agent spin $\mathbf s_i$};
        \draw[resultant] (0.55,0) -- (1.0,0);
        \node[key] at (1.05,0) {magnetisation $\mathbf m$};
    \end{scope}
\end{tikzpicture}

%% file: figs/factual-outcomes.tex
\begin{tikzpicture}[
    corr/.style={-{Latex[length=1.3mm]}, greenM, line width=0.7pt},
    hall/.style={-{Latex[length=1.3mm]}, redM, line width=0.7pt},
    disc/.style={draw=greyYed!55, fill=greyYed!7, line width=0.5pt},
    fork/.style={-{Latex[length=2.2mm]}, greyYed!75, line width=1.2pt},
    plab/.style={font=\footnotesize\bfseries, text=black},
    slab/.style={font=\footnotesize, text=greyYed}
]
    \def\R{0.52}
    \def\dx{3.0}
    \def\dy{0.8}

    \begin{scope}[shift={(0,0)}]
        \draw[disc] (0,0) circle (\R);
        \foreach \a in {20,110,200,290}{ \draw[corr] (0,0) -- (\a:\R); }
        \foreach \a in {65,155,245,335}{ \draw[hall] (0,0) -- (\a:\R); }
        \node[plab] at (0,-\R-0.30) {$t=0$: mixed};
        \node[slab] at (0,-\R-0.60) {$\rho\approx0.5,\ \bar c\approx0$};
    \end{scope}

    \begin{scope}[shift={(\dx,\dy)}]
        \draw[disc] (0,0) circle (\R);
        \foreach \a in {78,84,90,96,102}{ \draw[corr] (0,0) -- (\a:\R); }
        \node[plab, anchor=west] at (\R+0.14,0.10) {error correction};
        \node[slab, anchor=west] at (\R+0.14,-0.18) {$\rho\to0,\ \bar c\to1$};
    \end{scope}

    \begin{scope}[shift={(\dx,-\dy)}]
        \draw[disc] (0,0) circle (\R);
        \foreach \a in {78,84,90,96,102}{ \draw[hall] (0,0) -- (\a:\R); }
        \node[plab, anchor=west] at (\R+0.14,0.10) {shared hallucination};
        \node[slab, anchor=west] at (\R+0.14,-0.18) {$\rho\to1,\ \bar c\to1$};
    \end{scope}

    \draw[fork] (\R+0.06,0.12) to[out=40,in=195] ($(\dx,\dy)+(-\R-0.08,-0.06)$);
    \draw[fork] (\R+0.06,-0.12) to[out=-40,in=165] ($(\dx,-\dy)+(-\R-0.08,0.06)$);
\end{tikzpicture}